\documentclass[12pt]{iopart}

\usepackage{graphicx} 
\usepackage{amssymb}

\begin{document}

\title[Transport in exclusion processes with one-step memory]{Transport in
exclusion processes with one-step memory: density dependence and optimal
acceleration}

\author{Eial Teomy and Ralf Metzler}

\address{Institute for Physics \& Astronomy, University of Potsdam,
Karl-Liebknecht-Stra{\ss}e 24/25, D-14476 Potsdam-Golm, Germany}
\ead{eialteom@gmail.com}

\begin{abstract}
We study a lattice gas of persistent walkers, in which each site is occupied
by at most one particle and the direction each particle attempts to move to
depends on its last step. We analyse the mean squared displacement (MSD) of
the particles as a function of the particle density and their persistence
(the tendency to continue moving in the same direction).  For positive
persistence the MSD behaves as expected: it increases with the persistence
and decreases with the density. However, for strong anti-persistence we find
two different regimes, in which the dependence of the MSD on the density is
non-monotonic. For very strong anti-persistence there is an optimal density
at which the MSD reaches a maximum. In an intermediate regime, the MSD as a
function of the density exhibits both a minimum and a maximum, a phenomenon
which has not been observed before. We derive a mean-field theory which
qualitatively explains this behaviour.\\[0.2cm]
\emph{Keywords:} Exclusion process, persistence, lattice gas, memory, random
walk
\end{abstract}

\section{Introduction}

The active and passive motion of biological cells and the motion of their
internal components (molecular motors, enzymes, etc.) is a complicated
out-of-equilibrium process which occurs due to many factors, some of
them still unknown \cite{ref1}. This motion has been investigated
at the single-body level \cite{Hasnain2015,ref2,ref3}, many-body
level \cite{ref4,Berthier2013,ref5,Zimmermann2016,ref6,Reichhardt2014,
Zachreson2017,Reichhardt2014b,Lam2015,Graf2017,Illien2015}, or continuum
level \cite{ref7}. At the many-body level the focus is mostly on the
interactions between cells or bacteria, be they hydrodynamic \cite{ref4},
mutually aligning as in the Vicsek model \cite{ref5,Zimmermann2016},
energetic \cite{Zimmermann2016,ref6,Reichhardt2014,Zachreson2017}, or steric
\cite{Reichhardt2014b,Lam2015,Graf2017,Illien2015,Fisher2014}.

The motions of individual cells or bacteria are modelled in various ways, which
can be thought of as a random walk with a certain type of memory. One of the
most common models, motivated by experimental observations \cite{Berg1990},
is a run-and-tumble motion \cite{ref2,Reichhardt2014}, in which the walker
moves in a straight line for some time, and then abruptly changes its
direction. This model is captured by a memory term which gives a higher
probability of turning as more time passes from the last turn. A twitching
motion \cite{Zachreson2017} or motion with a self aligning director
\cite{Lam2015} is captured by a one-step memory term, i.e. the velocity at
each step depends on the velocity in the previous step but not on longer
memory terms. Other biological processes are also described as random walks
with memory \cite{Schulz2011,Ghosh2015b,Hermann2017}.

In random walks with memory, each step the walker makes depends not only
on its location in the previous step but on its history. It might depend
on its entire history, or a finite part of it. Notably, in one of the first
and best known random walk models that included memory \cite{Taylor1921},
a single walker moves on a one-dimensional lattice. At each step, the
walker either moves in the same direction as it did in the previous step
with probability $\frac{1}{2}+\delta$, or in the opposite direction with
probability $\frac{1}{2}-\delta$. This rule mimics inertia, and does not
introduce bias in any specific direction. The basic random walk model is
retrieved for $\delta=0$. Such walkers with one-step memory are also called
persistent walkers. Since the introduction of this model, it was expanded
in various forms to explain different phenomena in fields such as
polymer chains \cite{Tchen1952}, animal movement \cite{Kareiva1983}, scattering
in disordered media \cite{Boguna1998}, motion of bacteria \cite{Hasnain2015},
artificial micro-swimmers \cite{Romanczuk2012,Ghosh2015}, and motion in ordered
media \cite{Tahir}.

A different class of random walk models emulates the interactions in
many-body systems. In these "lattice-gas models" many walkers move on
a discrete graph or lattice with some type of interaction between the
different particles. In the Simple Symmetric Exclusion Principle (SSEP)
model \cite{Spitzer1970} the interaction is purely steric. Each site on a
lattice is either vacant or occupied by at most one walker, and each walker
has an internal clock, independent of the other walkers, which governs the
timing of its attempted moves. If a walker attempts to move to an already
occupied site, it remains in place. In the Asymmetric Simple Exclusion
Principle (ASEP) model \cite{Spitzer1970}, the walkers are biased to move in
a certain direction, and it has been used to describe transport phenomena in
biology \cite{Graf2017,Melbinger2011,vdz}. A special consideration is given
to one-dimensional systems \cite{ref8}, which emulate transport along a narrow
channel, such as transport of water \cite{Waghe2012} and drugs \cite{Yang2010}
through nanotubes, or of molecular motors in cellular protrusions
\cite{Graf2017} and along microtubules \cite{Melbinger2011,jae}. The
single file diffusion in one-dimensional systems is known to be anomalous,
even without memory \cite{Harris1965}. The basic SSEP and ASEP models have been
expanded to include energetic interactions \cite{Spohn1983}, a single biased
particle surrounded by unbiased particles \cite{ref10}, birth and death of
particles \cite{Markham2013}, higher site occupancy \cite{Arita2014}, spatial
inhomogeneities \cite{Szavitz2018} and kinetic constraints \cite{Ritort2003}.

There are several studies that combine these two variations of the basic
random walk, mutual exclusion effects and memory, and they investigate
three characteristics of this type of models. First, this model may be
considered as a coarse-grained version of active Brownian particles (ABP)
\cite{Romanczuk2012}, and it was shown that it indeed shows motility induced
phase separation \cite{Whitelam2017,gol}, one of the hallmarks of ABP. Second,
some studies derived an effective hydrodynamic description in either
one-dimensional \cite{Treloar2011,Kourbane2018} or higher-dimensional
\cite{Manacorda2017,Gavagnin2018} systems, including anomalous walkers
\cite{Arita2018}. The third group of studies investigates the mean squared
displacement (MSD) of crowded walkers with memory, in particular the short
time approximation of the MSD \cite{Galanti2013}, the MSD of interacting
subdiffusive random walkers in a one-dimensional system \cite{Sanders2014},
the MSD in the very high density limit in one-dimension \cite{Bertrand2018},
and the effective diffusion coefficient of a cross-shaped persistent walker
in a bath of memory-less cross-shaped walkers \cite{Chatterjee2018}.

In this study we investigate the MSD of persistent random walkers in a crowded
environment in both one-dimensional (1D) and two-dimensional (2D) systems. We
perform simulations covering the entire parameter space and find that in general
the MSD behaves as expected: it decreases with the density and increases
with the persistence. However, for strong anti-persistence we find two
different regimes, in which the dependence of the MSD on the density
is non-monotonic. For very strong anti-persistence there is an optimal
density at which the MSD reaches a maximum. In an intermediate regime,
the MSD as a function of the density exhibits both a minimum and a maximum,
a phenomenon which to our knowledge has not been observed before. We derive
a mean-field theory which explains this phenomenon qualitatively. We also
investigate the previously unexplored cases of totally persistent and totally
anti-persistent particles, for which the density has a critical effect. On
the one hand, a single totally persistent particle performs a ballistic motion,
while in a system with finite density all movement halts after a transient
time. On the other hand, a single totally anti-persistent walker is localised,
while in a system with density larger than $1/2$ its motion is unbounded.

The details of the model we investigate are described in section
\ref{sec_model}. Section \ref{sec_general} is devoted to the numerical
analysis of the MSD under general conditions, while in sections \ref{sec_tp}
and \ref{sec_tap} we consider the extreme cases of full persistence and
full anti-persistence. Finally, section \ref{sec_summary} summarises the
paper. In the appendix we derive a computationally efficient method to
calculate the MSD of a single particle with general memory, which we use in
our mean-field theory.

\section{Description of the model}
\label{sec_model}

We consider a lattice gas in either a 1D linear lattice or
a 2D square lattice. Each site on the lattice can be
either vacant or occupied by at most one particle. Each particle has an
independent exponential clock with mean time $\tau$. When the clock rings,
the particle attempts to move to one of its two (in 1D) or four (in 2D)
nearest neighbours. If the target site is vacant, the particle moves. Otherwise,
it remains in place. In both cases, its clock resets.

\begin{figure}
\centering
\includegraphics[width=0.6\columnwidth]{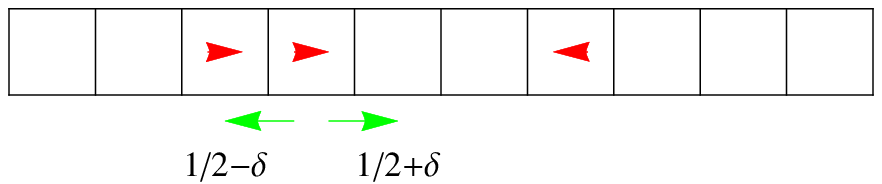}
\raisebox{-0.1\height}{\includegraphics[width=0.3\columnwidth]{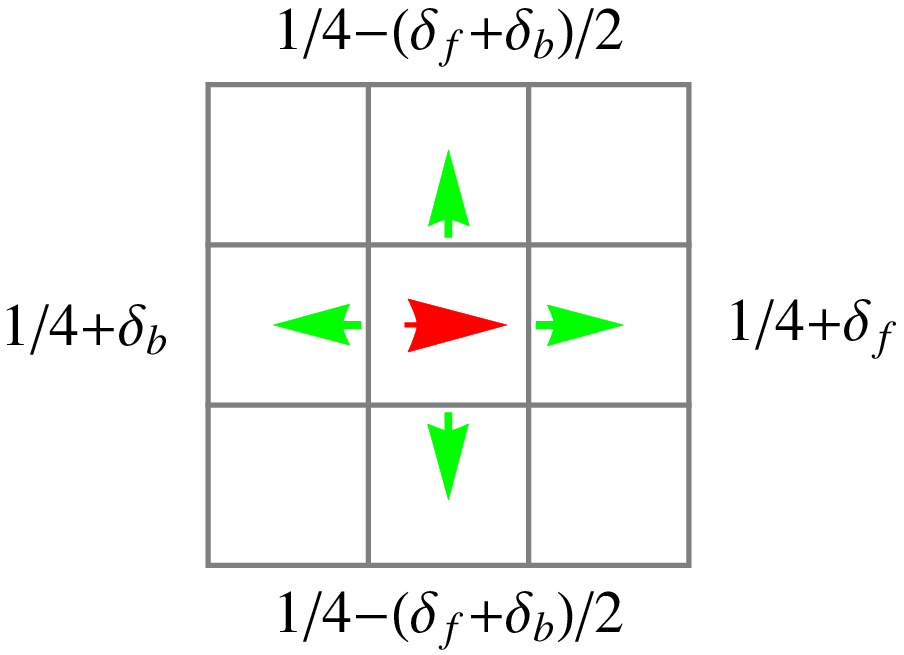}}
\caption{An illustration of the model. The last direction in which the particle
moved is denoted by the red arrow. At each step the particle turns in one
of the directions with probabilities shown near the green arrows, and moves
in that direction if the target site is vacant. In 1D the system parameter is the
persistence $\delta$, in 2D we classify the motion by the two parameters $\delta_f$
and $\delta_b$.}
\label{model}
\end{figure}

The target direction, however, is not chosen from a uniform distribution but
it rather depends on the history of the particle. We consider here one-step
memory models, also called persistent walkers. In the 1D model,
pictured in figure \ref{model}a, the probability that a particle attempts to
move in the same direction as in its previous state is $\frac{1}{2}+\delta$
with $-\frac{1}{2}\leq\delta\leq\frac{1}{2}$, and the probability it reverses
its direction is $\frac{1}{2}-\delta$. We call the parameter $\delta$ the
persistence parameter, since it encodes the tendency of the particle to persist
in its motion. Note that since the probability distribution for choosing
the direction of motion is relative to the current direction of motion,
there is no macroscopic bias in the system unless it is imposed from the
boundaries. In the 2D model, pictured in figure \ref{model}b, the
probability that a particle attempts to move in the same direction as
before is $\frac{1}{4}+\delta_{f}$, the probability that it attempts
to move in the opposite direction is $\frac{1}{4}+\delta_{b}$, and the
probability that it attempts to move in either of the other two directions
is $\frac{1}{4}-\frac{\delta_{f}+\delta_{b}}{2}$, i.e. here $\delta_{f}$
is not necessarily equal to $-\delta_{b}$ as in one dimension.
\footnote{In higher dimensions the situation is similar to 2D:
the probability to move forward is $\frac{1}{2d}+\delta_{f}$,
the probability to move backwards is $\frac{1}{2d}+\delta_{b}$,
and the probability to move to any of the other $2d-2$ directions is
$\frac{1}{2d}-\frac{\delta_{f}+\delta_{b}}{2\left(d-1\right)}$. We expect the
results for high dimensions to be qualitatively similar to two dimensions.
Here we focus on the relevant 1D and 2D cases.}

Note that although the net current is zero, this model is out of equilibrium
because it does not obey detailed balance. Consider for example a particle
moving to the vacant site to its right, and that in its previous step it also
moved to the right. Such a move occurs with probability $\frac{1}{2}+\delta$
(in 1D) or $\frac{1}{4}+\delta_{f}$ (in 2D). The opposite transition,
however, has a zero probability of occurring, since if the particle moves
to the now vacant adjacent site to its left its last move was to the left,
and it is thus in a different state than the one it started from.

In the simulations we perform, we use periodic boundary conditions and
a system of either size $10^{4}$ (in 1D) or $100\times100$ (in 2D). The
initial state of the system is uncorrelated: each site is independently
occupied with probability $\rho$, and the direction of each walker is
independently and uniformly generated from the possible two (in 1D) or four
(in 2D) directions. At each step of the simulation, one of the particles is
chosen randomly and a move is attempted. Whether the move succeeds or not
the clock advances by $\frac{\tau}{N}$, where $N$ is the number of particles
in the system and we choose the time units to be $\tau=1$. This evolution is
equivalent to each particle having an exponential clock with mean $\tau$. All
results are averages of $100$ independent runs.

\section{Finite persistence}
\label{sec_general}

\begin{figure}
\centering
\includegraphics[width=5.0cm]{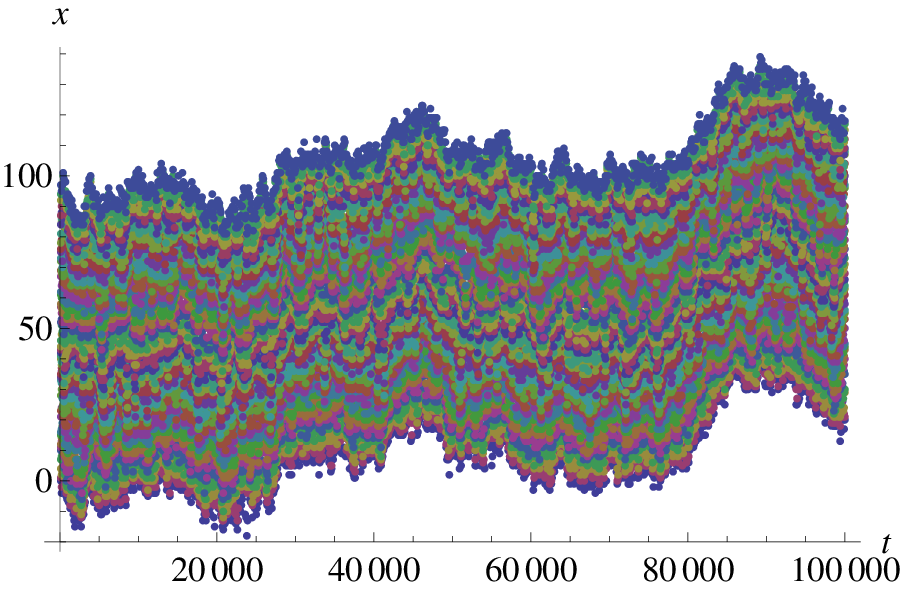}
\includegraphics[width=5.0cm]{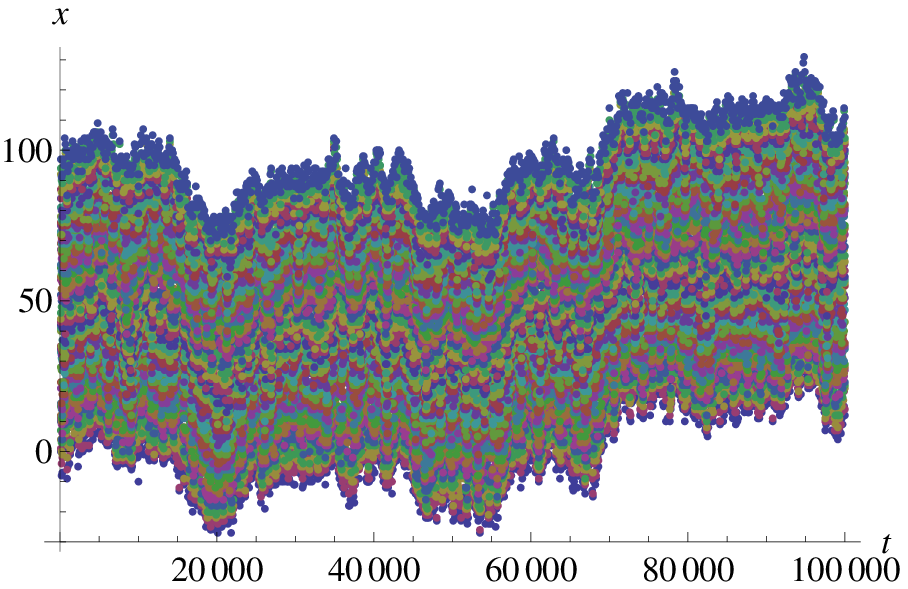}
\includegraphics[width=5.0cm]{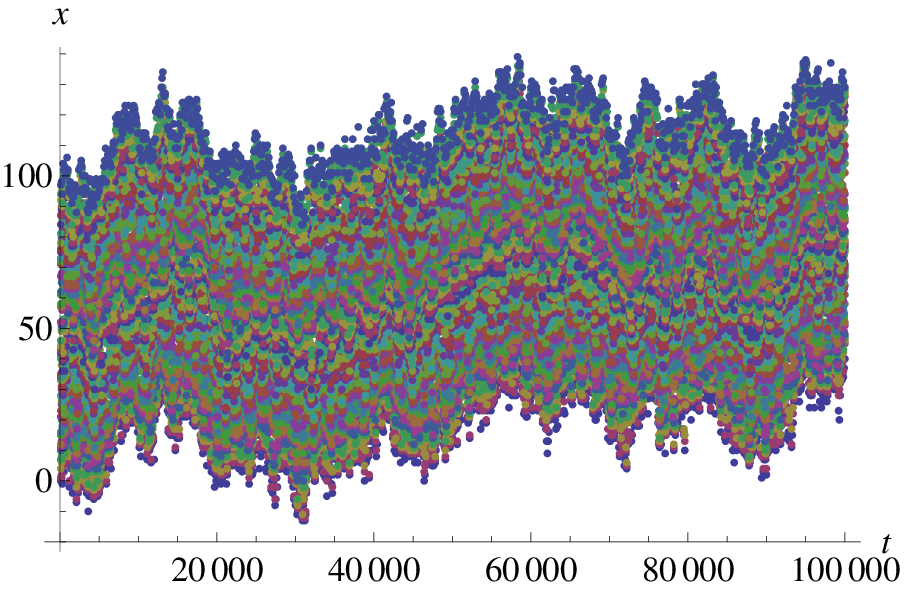}
\caption{Sample trajectories of 1D systems with density $\rho=0.4$. Left:
anti-persistent case, $\delta=-0.2$; Middle: neutral case, $\delta=0$;
Right: persistent case, $\delta=0.2$.}
\label{trajs}
\end{figure}

\begin{figure}
\centering
\includegraphics[height=4.0cm]{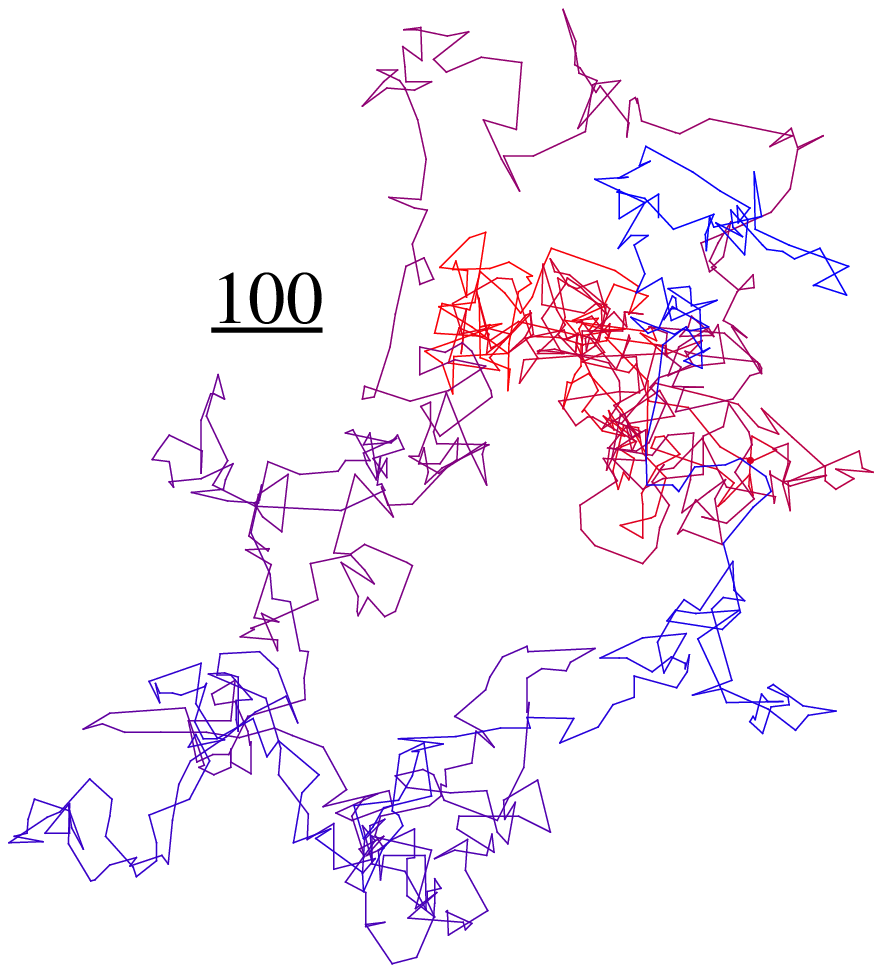}
\includegraphics[height=4.0cm]{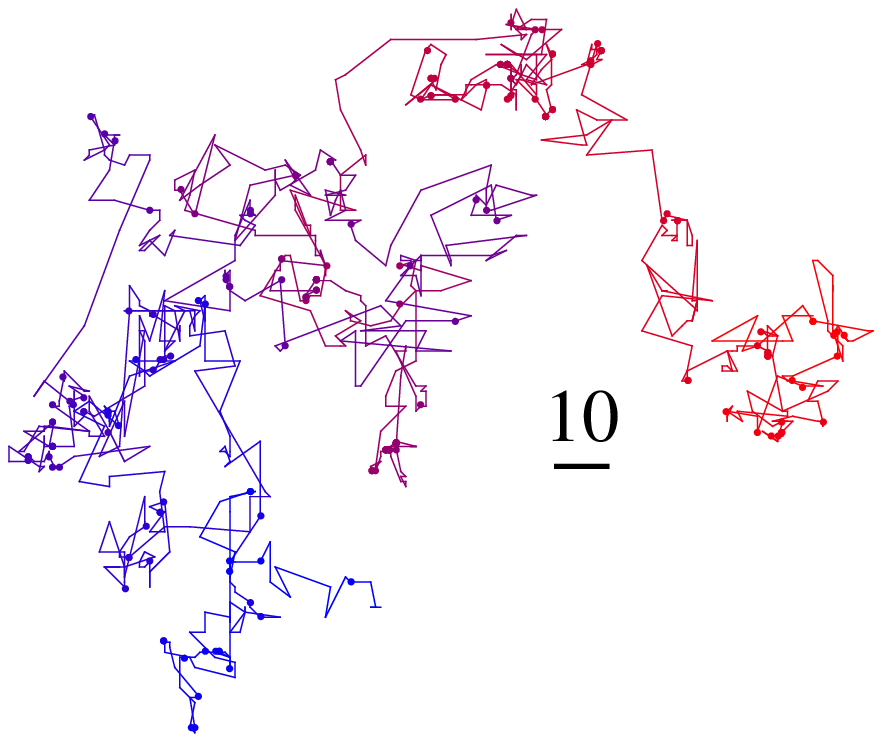}
\includegraphics[height=4.0cm]{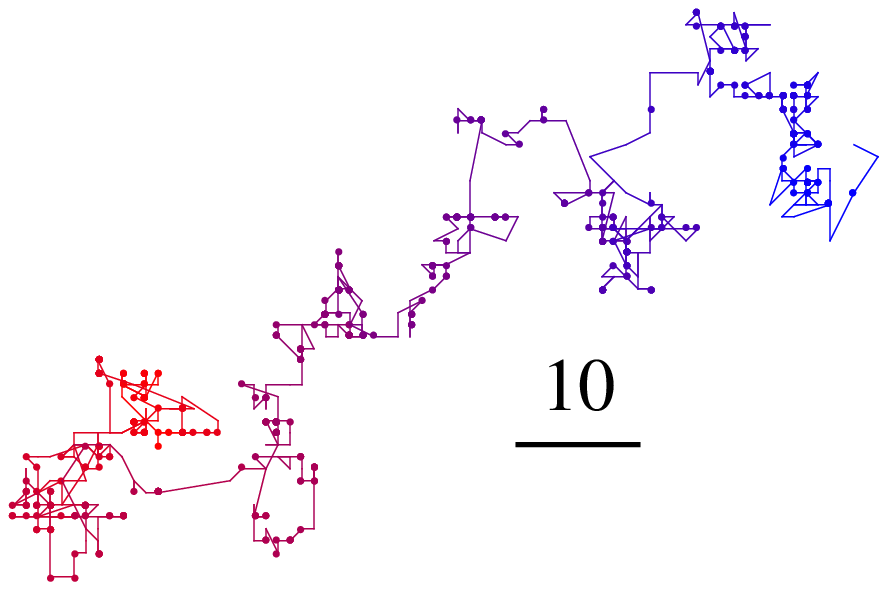}\\[0.4cm]
\includegraphics[height=2.76cm]{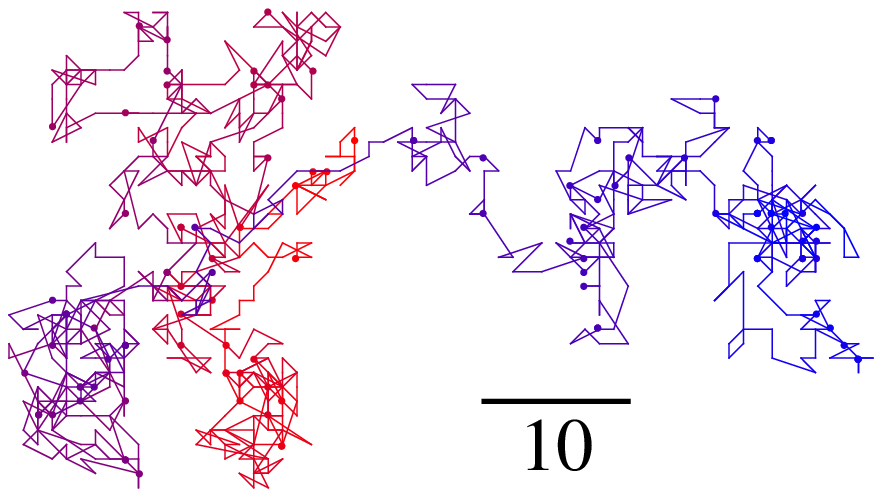}
\includegraphics[height=2.76cm]{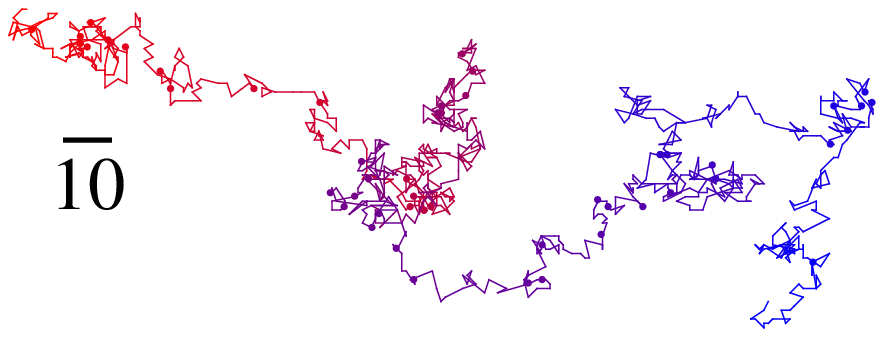}
\includegraphics[height=2.76cm]{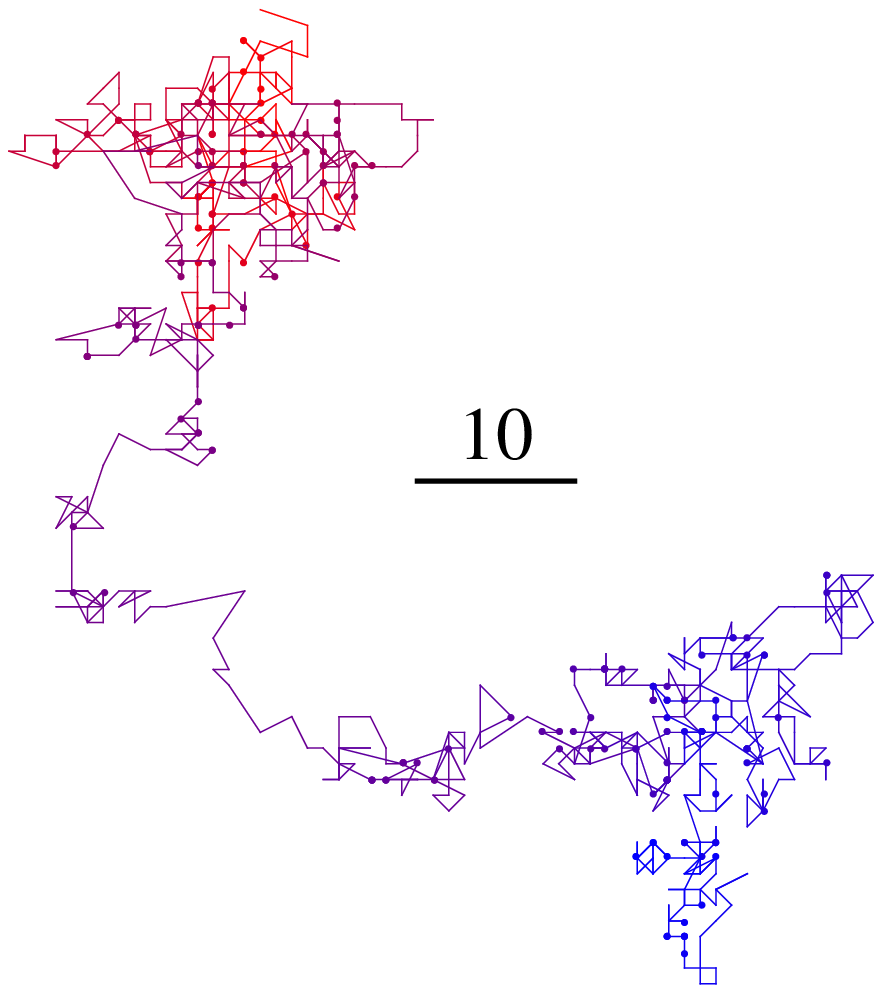}
\caption{Sample trajectories of 2D systems for densities $\rho=0.1$ (Left),
$\rho=0.6$ (Middle) and $\rho=0.9$ (Right). In the top row the persistence
parameters are $\delta_f=0.45$ and $\delta_b=-0.50$ (highly persistent case),
in the bottom row they are $\delta_f=-0.50$ and $\delta_b=0.45$ (highly
antipersistent case). The particle starts in the red region and ends in the
blue region. The lengths of the black scale bars is given in lattice units.
The different sizes of the scale bars give an impression of the overall
span of the trajectories with respect to each other.}
\label{trajs2d}
\end{figure}

We first consider the non-pathological cases, i.e., that neither the
probability to continue forward nor the probability to turn backward is
exactly unity---we will consider these limiting cases below. Sample
trajectories for three cases are shown in figure \ref{trajs}. Figure
\ref{trajs2d} shows sample trajectories in 2D, here we only depict a
single particle's trajectory for clarity.

\begin{figure}
\centering
\includegraphics[height=4.8cm]{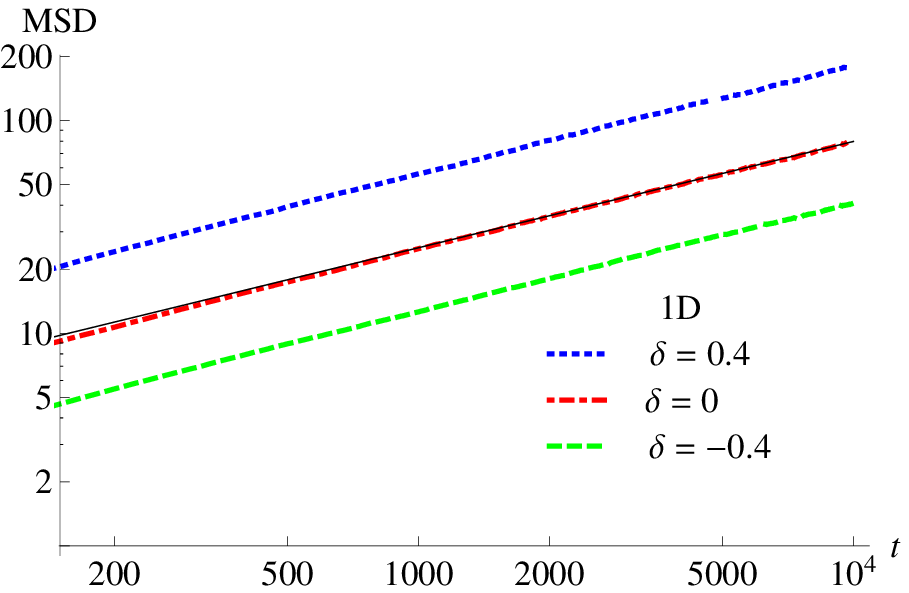}
\includegraphics[height=4.8cm]{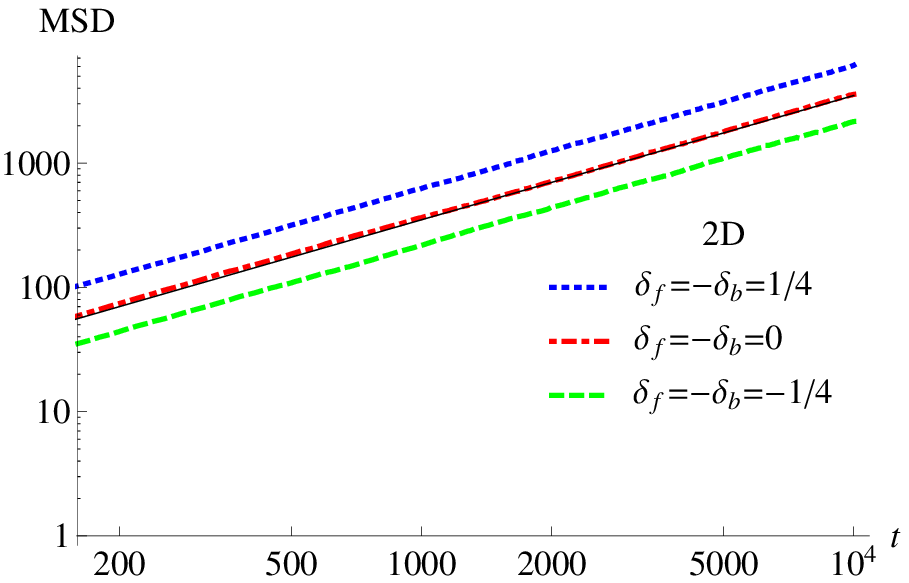}
\caption{The MSD for a 1D and a 2D system with
$\rho=0.5$ and different persistence values. The continuous black line is the known
result for $\delta=0$ in 1D, equation (\ref{msd1dnomem}), or a numerical fit
in 2D. In all cases, the MSD grows with time as $\sqrt{t}$ for 1D and as $t$
for 2D.}
\label{fig_msd}
\end{figure}

Without memory, it is known that in one dimensional systems
the MSD scales as $\sqrt{t}$ \cite{Harris1965}, with the dependence on the
density, for an equilibrium initial condition, given by
\begin{equation}
\label{msd1dnomem}
\left\langle x^2\right\rangle_{1D}(t)=\frac{1-\rho}{\rho}\sqrt{\frac{2D_0t}{\pi}},
\end{equation}
where $D_{0}$ is the diffusion coefficient of a single walker. Note
that for a single particle the MSD is linear in time $\left\langle
x^{2}\right\rangle_{1D}(t)=D_{0}t$ so that the limit $\rho\to0$ drives the
system across a transition. In two and higher dimensions, the MSD
grows linearly with $t$, but the exact coefficient is unknown analytically
\cite{Arita2014,ref9}. Furthermore, the MSD of a single particle with
finite memory grows linearly with time in any dimension, since the velocity
correlations decay exponentially. For one-step memory, the MSD of a single
particle is \cite{Taylor1921,Tchen1952}
\begin{eqnarray}
\langle x^2\rangle_{1D}(t)=D_0\frac{1+2\delta}{1-2\delta}t,\nonumber\\
\langle\mathbf{r}^2\rangle_{2D}(t)=2D_0\frac{1+\delta_f-\delta_b}{1-\delta_f
+\delta_b}t ,
\end{eqnarray}
where $D_0$ is the diffusion coefficient of a single walker without memory, and
$\mathbf{r}^2=x^2+y^2$. Except for the pathological cases of full persistence
and full anti-persistence, we find numerically that for all values of the
persistence $\delta$ and the density $\rho$ the MSD in 1D systems grows as
$\sqrt{t}$ and in 2D systems as $t$, see figure \ref{fig_msd}. The slope,
however, becomes distinctly different from memory-less systems.

\begin{figure}
\centering
\includegraphics[height=4.8cm]{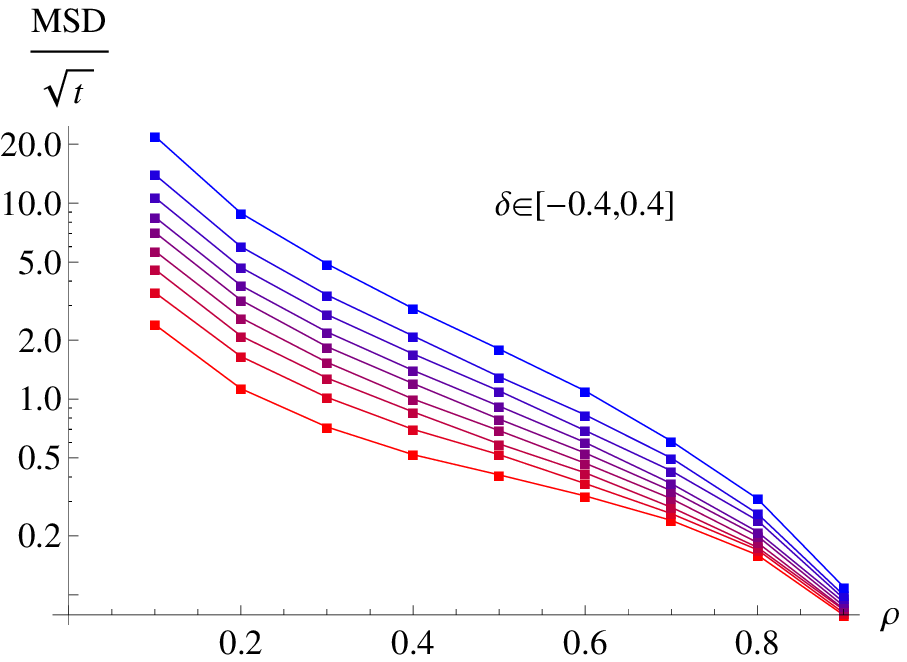}
\includegraphics[height=4.8cm]{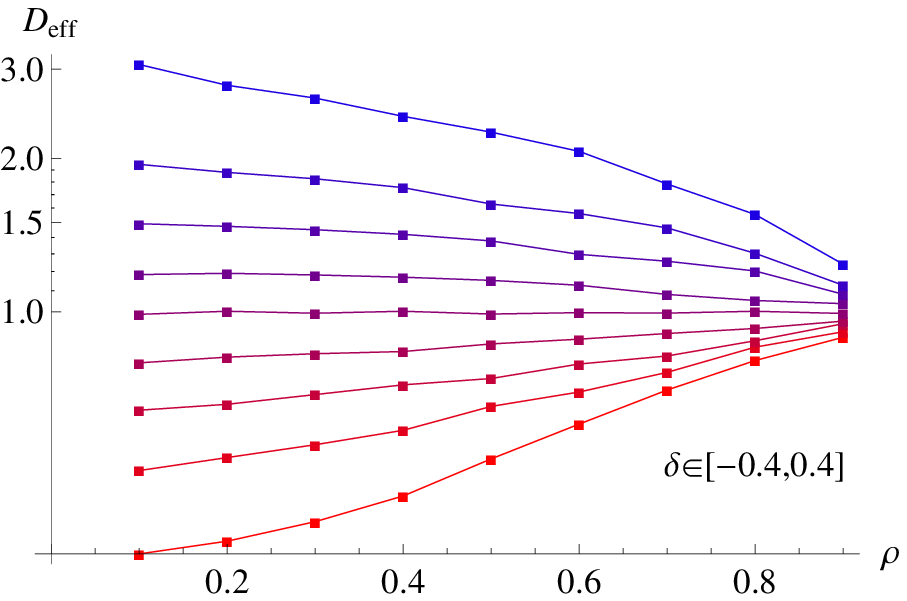}\\
\includegraphics[height=4.8cm]{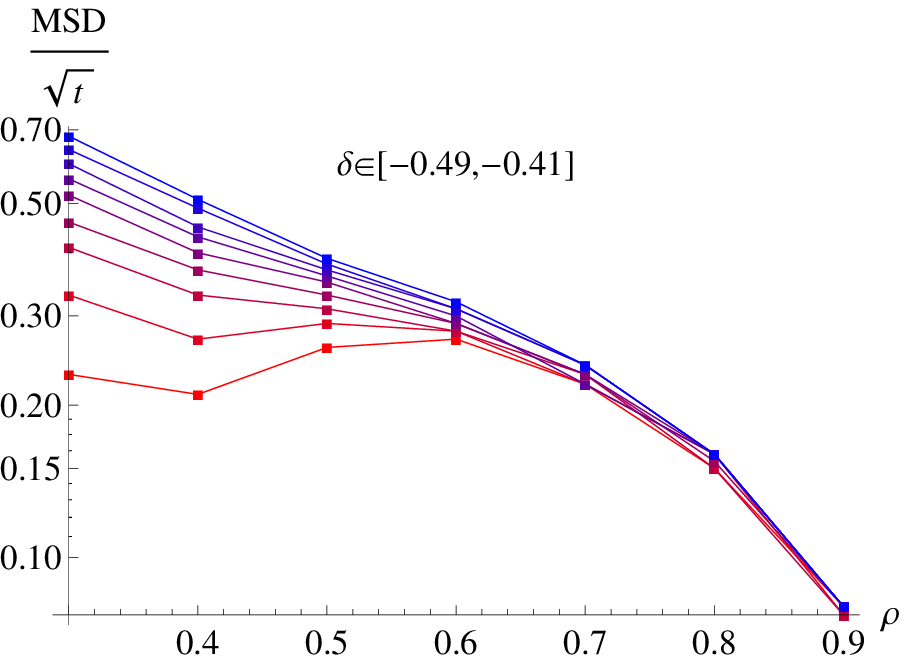}
\includegraphics[height=4.8cm]{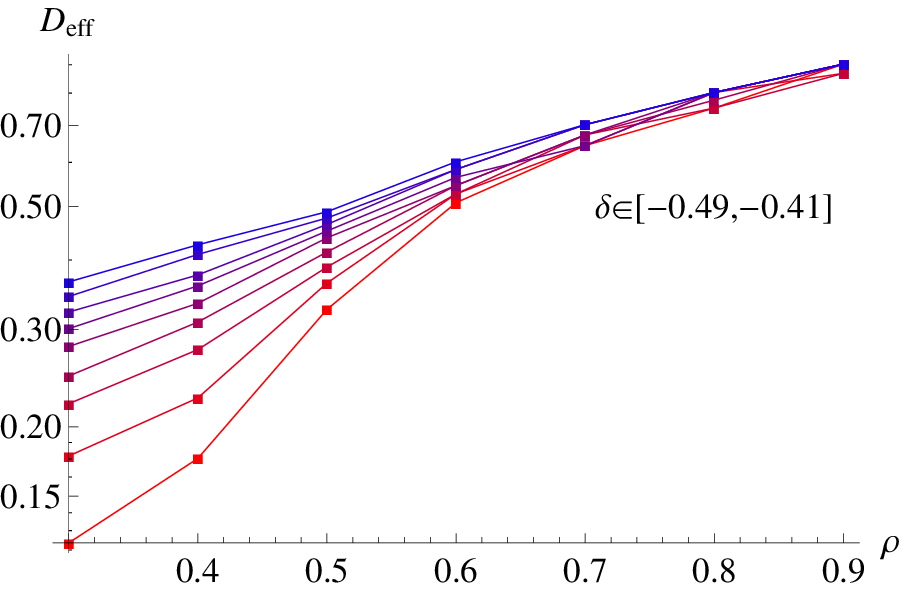}\\
\caption{Log-plots of the slope of the MSD curve in 1D and the extracted
diffusion coefficient as a function of the density $\rho$ for different
values of the persistence $\delta$ from $\delta=0.4$ (top line in blue)
to $\delta=-0.4$ (bottom line in red) in jumps of $0.1$ in the top panels,
and from $\delta=-0.49$ (top line in blue) to $\delta=-0.41$ (bottom line
in red) in jumps of $0.01$ in the bottom panels.}
\label{fig_deff}
\end{figure}

\begin{figure}
\centering
\includegraphics[height=4.8cm]{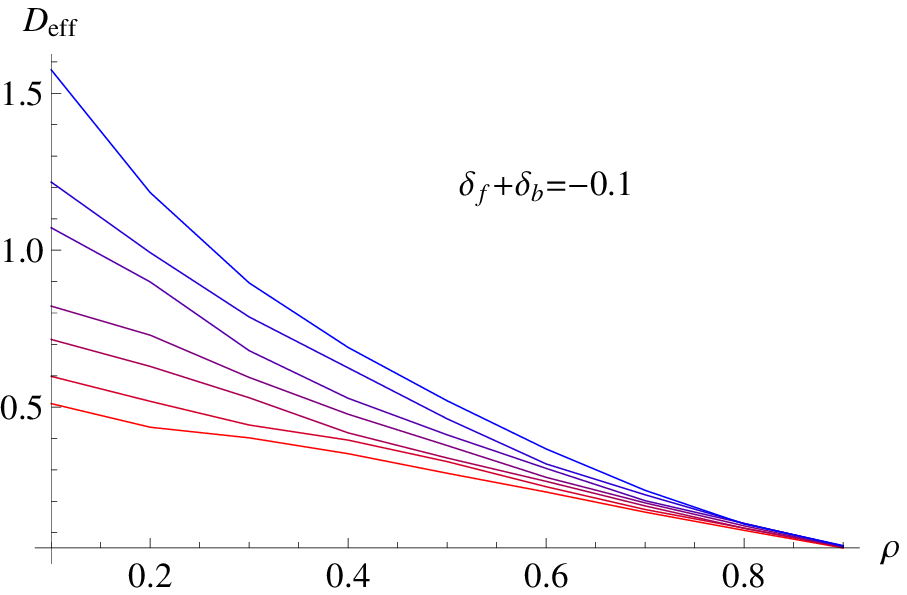}
\includegraphics[height=4.8cm]{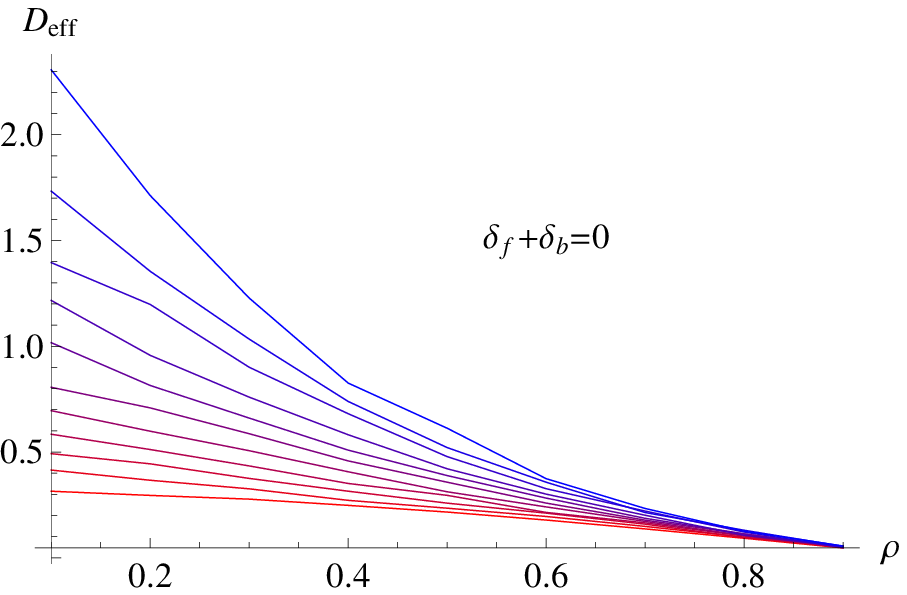}\\
\includegraphics[height=4.8cm]{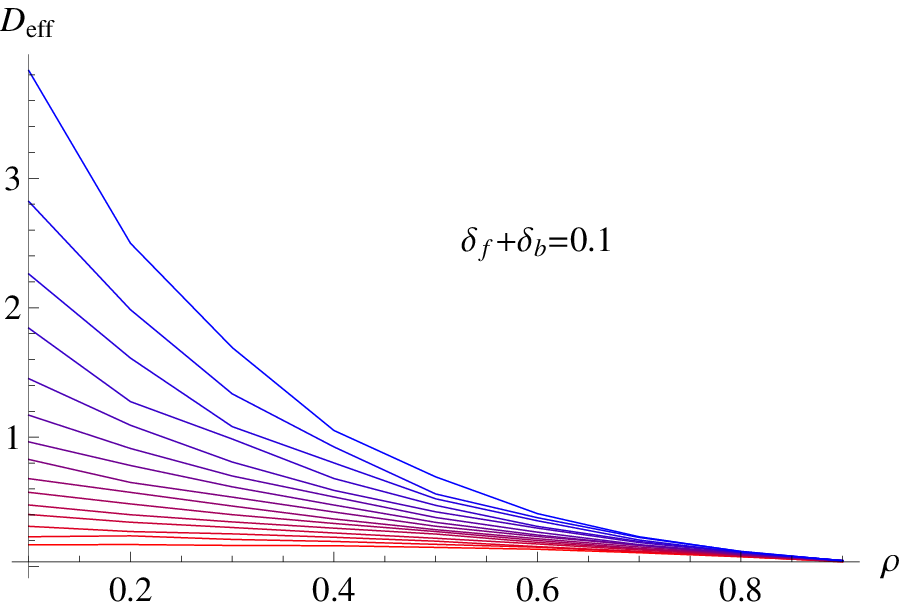}
\includegraphics[height=4.8cm]{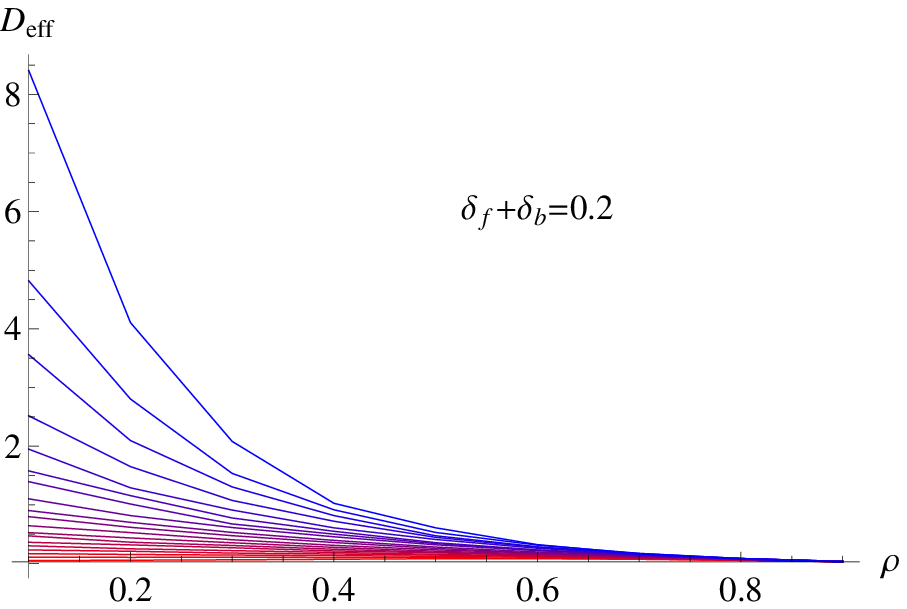}
\caption{Diffusion coefficient for the 2D one-step
memory model as a function of the density $\rho$ for various values of
$\delta_{f}$ and $\delta_{b}$ and fixed $\delta_{f}+\delta_{b}$. The
different lines correspond to different values of $\delta_{f}$
and $\delta_{b}$. In each plot the top line (in blue) is for
$\delta_{b}=-0.25$ and the bottom line (in red) is for $\delta_{f}=-0.25$. For
example, the lines for $\delta_{f}+\delta_{b}=-0.1$ are (from top to bottom)
$(\delta_{b}=-0.25,\delta_{f}=0.35),(\delta_{b}=-0.2,\delta_{f}=0.3),...,(\delta_{b}=0.35,\delta_{f}=-0.25)$.
The limit at $\rho\rightarrow0$ agrees with the single particle values.}
\label{msd2D_fig}
\end{figure}

In 1D for each combination of the density $\rho$ and the persistence
$\delta$ we use equation (\ref{msd1dnomem}) and extract an effective diffusion
coefficient $D_{\mathrm{eff}}$, as shown in figure \ref{fig_deff}, while
in 2D we extract an effective diffusion coefficient from $\left\langle
\mathbf{r}^2\right\rangle=2D_{\mathrm{eff}}t$, see figure \ref{msd2D_fig}. Since
for memory-less systems the MSD is a decreasing function of the density,
and for single persistent walkers the MSD is an increasing function of the
persistence, we expect that this dependence remains when both density
and persistence are involved. Indeed, in 1D, we find that the effective
diffusion coefficient is always a monotonically decreasing function of the
persistence $\delta$, while in 2D it is a monotonically decreasing function
of $\delta_{f}$ and a monotonically increasing function of $\delta_{b}$, as
intuitively expected. Furthermore, in 1D it is an increasing function of the
density for $\delta<0$ and a decreasing function of the density for $\delta>0$.

\begin{figure}
\centering
\includegraphics[height=3.6cm]{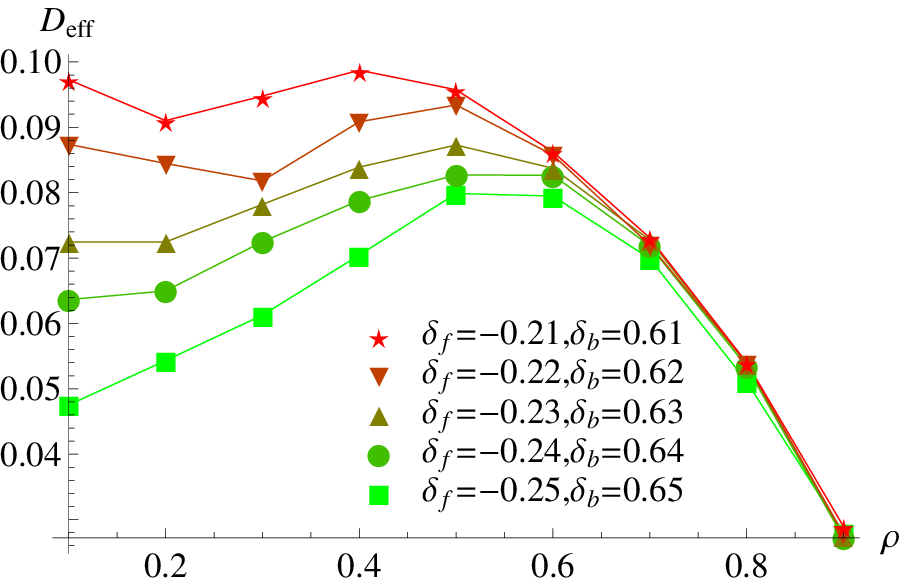}
\includegraphics[height=3.6cm]{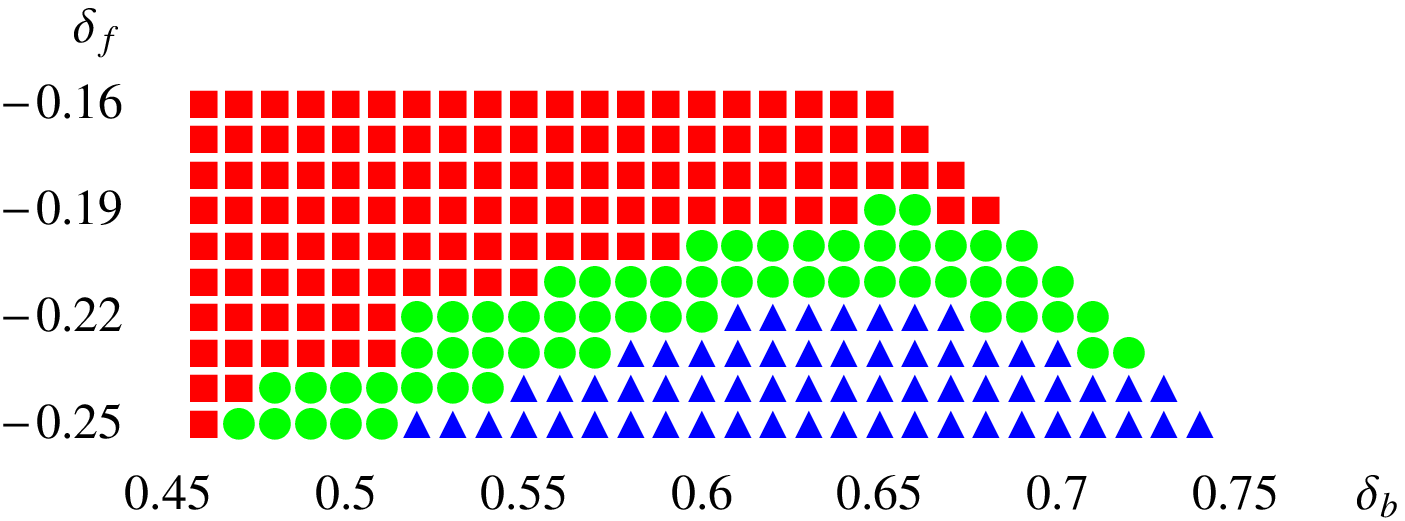}
\caption{(a) Diffusion coefficient for the 2D one-step memory model as
a function of the density $\rho$ for various values of $\delta_f$ and
$\delta_b$. The connecting lines are a guide to the eye. When $\delta_{f}$
is small and $\delta_b$ is large, $D$ is not monotonic with $\rho$. (b)
Phase diagram in the $\delta_b-\delta_f$ plane showing for each value whether
the MSD is monotonically decreasing with $\rho$ (red squares), has a single
maximum (blue triangles), or has both a maximum and a minimum (green circles).}
\label{msd2D_anti_fig}
\end{figure}

We now look at the dependence of $\langle x^2\rangle/\sqrt{t}$ in 1D and of
$\langle\mathbf{r}^2\rangle/t$ in 2D on the density. In most cases, it is
a decreasing function of the density. However, at strong anti-persistence
($\delta\leq-0.48$ in 1D and $\delta_b\geq0.55$ and $\delta_f\leq-0.19$
in 2D), remarkably we find two other types of density dependence. In that
regime, the MSD may either have a single maximum as a function of the
density (for example, as in $\delta_f=-0.25$ and $\delta_b=0.65$),
or it may have both a maximum and a minimum (as in $\delta_f=-0.22$ and
$\delta_b=0.62$), see figure \ref{msd2D_anti_fig}. We note that in 1D,
the single maximum regime occurs only for totally anti-persistent particles
($\delta=-1/2$), as will be explained below.

A similar peak in the MSD as a function of density was also found in a
lattice model of a single biased tracer surrounded by regular random walkers
\cite{Illien2015,Illien2018}, but to our knowledge no other model exhibits the
more complicated behaviour of both a maximum and a minimum. This behaviour is
a competition between three mechanisms. The first, simplest mechanism occurs
at high densities and is simply the blocking of the movement of the particles
by their neighbours which reduces the MSD. The second mechanism is different
in 1D systems and in higher dimensional system, but in both cases it occurs
at low densities. In 1D the MSD of a single particle scales linearly with $t$
while the MSD in a system with finite density scales as $\sqrt{t}$ due to the
known effects of single file diffusion \cite{Harris1965,Lizana2010}. Therefore,
the MSD divided by $\sqrt{t}$ diverges at $\rho=0$, and thus for very small
densities it is a decreasing function of $\rho$. In higher dimensions,
the explanation is different since both the single particle MSD and the
MSD in a finite density system are linear in time. In this case, when the
anti-persistence is not too high and the density is low, the collisions
between the particles are rare, and thus the only effect of the density is
to occasionally block the movement, and thus reduce the MSD.

The intriguing third mechanism becomes relevant at high
degrees of anti-persistence. In this mechanism, a particle is prevented
from moving backwards by other particles that reach the site it occupied
before, and thus in effect force it to move forward. In order to explain
this behaviour qualitatively, we consider a single tracer starting at the
origin. The other particles are acting as an effective bath, such that each
move succeeds with probability $1-\rho$ and fails with probability $\rho$. We
consider this mean-field description in 1D and 2D.

In 1D the time evolution of the probability to find the
particle at time $t$ in site $n$ such that in the last step it moved in
direction $\sigma=\pm1$, $Q_{\sigma}(n,t)$, is
\begin{eqnarray}
\tau\frac{\partial Q_{\sigma}(n,t)}{\partial t}&=&-Q_{\sigma}(n,t)+\left(
\frac{1}{2}+\delta\right)\left[\rho Q_{\sigma}(n,t)+\left(1-\rho\right)Q_{
\sigma}(n-\sigma,t)\right]\nonumber\\
&&+\left(\frac{1}{2}-\delta\right)\left[\rho Q_{-\sigma}(n,t)+\left(1-\rho
\right)Q_{-\sigma}(n-\sigma,t)\right] .
\end{eqnarray}
Using the discrete Fourier transform of $Q_{\sigma}(n,t)$
\begin{equation}
\tilde{Q}_{\sigma}(k,t)=\sum^{\infty}_{n=-\infty}e^{ink}Q_{\sigma}(n,t) ,
\end{equation}
yields
\begin{eqnarray}
\tau\frac{\partial\tilde{Q}_{\sigma}(k,t)}{\partial t}&=&-\tilde{Q}_{\sigma}
(k,t)+\left(\frac{1}{2}+\delta\right)\left[\rho \tilde{Q}_{\sigma}(k,t)+\left(
1-\rho\right)e^{i\sigma k}\tilde{Q}_{\sigma}(k,t)\right]\nonumber\\
&&+\left(\frac{1}{2}-\delta\right)\left[\rho \tilde{Q}_{-\sigma}(k,t)+\left(1
-\rho\right)e^{i\sigma k}Q_{-\sigma}(k,t)\right] .
\end{eqnarray}
In matrix form this may be written as
\begin{equation}
\tau\frac{\partial}{\partial t}\left(\begin{array}{c}\tilde{Q}_{+}\\\tilde{Q}_{-}\end{array}\right)={\cal M}_{1}(k)\left(\begin{array}{c}\tilde{Q}_{+}\\\tilde{Q}_{-}\end{array}\right) ,
\end{equation}
with
\begin{equation}
\fl{\cal M}_{1}(k)=\left(\begin{array}{cc}-1+\left(\frac{1}{2}+\delta\right)\left[\rho+\left(1-\rho\right)e^{ik}\right]&\left(\frac{1}{2}-\delta\right)\left[\rho+\left(1-\rho\right)e^{ik}\right]\\
\left(\frac{1}{2}-\delta\right)\left[\rho+\left(1-\rho\right)e^{-ik}\right]&-1+\left(\frac{1}{2}+\delta\right)\left[\rho+\left(1-\rho\right)e^{-ik}\right]\end{array}\right) .
\end{equation}
Therefore
\begin{equation}
\left(\begin{array}{c}\tilde{Q}_{+}(k,t)\\\tilde{Q}_{-}(k,t)\end{array}\right)=e^{{\cal M}_{1}(k)t/\tau}\left(\begin{array}{c}\tilde{Q}_{+}(k,0)\\\tilde{Q}_{-}(k,0)\end{array}\right) .
\end{equation}
In order to find $\tilde{Q}_{\sigma}(k,0)$ we note that by its definition
\begin{equation}
\tilde{Q}_{\sigma}(k,0)=\sum^{\infty}_{n=-\infty}e^{ink}Q_{\sigma}(n,0)=\sum^{\infty}_{n=-\infty}e^{ink}\frac{1}{2}\delta_{n,0}=\frac{1}{2} .
\end{equation}
The MSD is given by
\begin{eqnarray}
\left\langle n^2\right\rangle&=&-\left.\frac{\partial^2}{\partial k^2}\left[\tilde{
Q}_{+}(k,t)+\tilde{Q}_{-}(k,t)\right]\right|_{k=0}\nonumber\\
&=&\frac{4\delta\left(1-\rho\right)^{2}}{\left(1-2\delta\right)^{2}}\left(e^{-(1-2
\delta)t/\tau}-1\right)+\left(1-\rho\right)\left(1+\frac{4\delta\left(1-\rho\right)}{
1-2\delta}\right)\frac{t}{\tau}.
\label{first}
\end{eqnarray}
The long time MSD is monotonically decreasing with the density for all
$\delta\geq-\frac{1}{6}$, while for $\delta<-\frac{1}{6}$ it has a single
maximum at $\rho=\frac{1+6\delta}{8\delta}$.

We now consider a persistent walker in 2D, such that each move succeeds
with probability $1-\rho$ and fails with probability $\rho$. The evolution
equation for the probability to find the walker at site $\mathbf{n}$ at
time $t$, such that in the last step it moved in direction $\mathbf{e}$,
$Q_{\mathbf{e}}\left(\mathbf{n},t\right)$ is
\begin{eqnarray}
\fl\tau\frac{\partial Q_{\mathbf{e}}\left(\mathbf{n},t\right)}{\partial t}&=&-Q_{
\mathbf{e}}\left(\mathbf{n},t\right)+\left(\frac{1}{4}+\delta_{f}\right)\left[
\rho Q_{\mathbf{e}}\left(\mathbf{n},t\right)+\left(1-\rho\right)Q_{\mathbf{e}}
\left(\mathbf{n}-\mathbf{e},t\right)\right]\nonumber\\
\fl&&+\left(\frac{1}{4}+\delta_{b}\right)\left[\rho Q_{-\mathbf{e}}\left(\mathbf{n},
t\right)+\left(1-\rho\right)Q_{-\mathbf{e}}\left(\mathbf{n}-\mathbf{e},t\right)
\right]\nonumber\\
\fl&&+\left(\frac{1-2(\delta_{f}+\delta_{b})}{4}\right)\sum_{s=\pm1}\left[\rho Q_{
s\mathbf{e}^{\perp}}\left(\mathbf{n},t\right)+\left(1-\rho\right)Q_{s\mathbf{e}^{
\perp}}\left(\mathbf{n}-\mathbf{e},t\right)\right] ,
\end{eqnarray}
where $\mathbf{e}^{\perp}=\mathbf{e}_{y}$ if $\mathbf{e}=\pm\mathbf{e}_{x}$, and $\mathbf{e}^{\perp}=\mathbf{e}_{x}$ if $\mathbf{e}=\pm\mathbf{e}_{y}$. Using the Fourier transform of $Q_{\mathbf{e}}\left(\mathbf{n},t\right)$
\begin{equation}
\tilde{Q}_{\mathbf{e}}\left(\mathbf{k},t\right)=\sum^{\infty}_{n_{x},n_{y}=-\infty}e^{i\mathbf{n}\cdot\mathbf{k}}Q_{\mathbf{e}}\left(\mathbf{n},t\right) ,
\end{equation}
yields
\begin{eqnarray}
\fl\tau\frac{\partial \tilde{Q}_{\mathbf{e}}\left(\mathbf{k},t\right)}{\partial t}
&=&-\tilde{Q}_{\mathbf{e}}\left(\mathbf{k},t\right)+\left(\frac{1}{4}+\delta_{f}
\right)\left[\rho \tilde{Q}_{\mathbf{e}}\left(\mathbf{k},t\right)+\left(1-\rho
\right)e^{i\mathbf{e}\cdot\mathbf{k}}\tilde{Q}_{\mathbf{e}}\left(\mathbf{k},t
\right)\right]\nonumber\\
\fl&&+\left(\frac{1}{4}+\delta_{b}\right)\left[\rho \tilde{Q}_{-\mathbf{e}}\left(
\mathbf{k},t\right)+\left(1-\rho\right)e^{i\mathbf{e}\cdot\mathbf{k}}Q_{-\mathbf{
e}}\left(\mathbf{k},t\right)\right]\nonumber\\
\fl&&+\left(\frac{1-2(\delta_{f}+\delta_{b})}{4}\right)\sum_{s=\pm1}\left[\rho
\tilde{Q}_{s\mathbf{e}^{\perp}}\left(\mathbf{k},t\right)+\left(1-\rho\right)e^{
i\mathbf{e}\cdot\mathbf{k}}Q_{s\mathbf{e}^{\perp}}\left(\mathbf{k},t\right)\right] .
\end{eqnarray}
In matrix form this may be written as
\begin{equation}
\tau\frac{\partial}{\partial t}\left(\begin{array}{c}\tilde{Q}_{x}\\\tilde{Q}_{-x}\\\tilde{Q}_{y}\\\tilde{Q}_{-y}\end{array}\right)={\cal M}_{2}(k)\left(\begin{array}{c}\tilde{Q}_{x}\\\tilde{Q}_{-x}\\\tilde{Q}_{y}\\\tilde{Q}_{-y}\end{array}\right) .
\end{equation}
Therefore
\begin{equation}
\left(\begin{array}{c}\tilde{Q}_{x}\left(\mathbf{k},t\right)\\\tilde{Q}_{-x}\left(\mathbf{k},t\right)\\\tilde{Q}_{y}\left(\mathbf{k},t\right)\\\tilde{Q}_{-y}\left(\mathbf{k},t\right)\end{array}\right)=e^{{\cal M}_{2}(k)t/\tau}\left(\begin{array}{c}\tilde{Q}_{x}\left(\mathbf{k},0\right)\\\tilde{Q}_{-x}\left(\mathbf{k},0\right)\\\tilde{Q}_{y}\left(\mathbf{k},0\right)\\\tilde{Q}_{-y}\left(\mathbf{k},0\right)\end{array}\right) .
\end{equation}
In order to find $\tilde{Q}_{\mathbf{e}}\left(\mathbf{k},0\right)$ we note that by its definition
\begin{equation}
\tilde{Q}_{\mathbf{e}}\left(\mathbf{k},0\right)=\sum^{\infty}_{n_{x},n_{y}=-\infty}e^{i\mathbf{n}\cdot\mathbf{k}}Q_{\mathbf{e}}\left(\mathbf{n},0\right)=\sum^{\infty}_{n_{x},n_{y}=-\infty}e^{i\mathbf{n}\cdot\mathbf{k}}\frac{1}{4}\delta_{\mathbf{n},0}=\frac{1}{4} .
\end{equation}
The long time limit of the MSD is given by
\begin{eqnarray}
\nonumber
\left\langle\mathbf{n}^2\right\rangle&=&-\left.\mathbf{\nabla}^{2}_{k}\sum_{
\mathbf{e}}\tilde{Q}_{\mathbf{e}}\left(\mathbf{k},t\right)\right|_{k=0}\\
&=&\frac{1-\rho}{1+\delta_{b}-\delta_{f}}\left[1+\left(1-2\rho\right)\left(
\delta_f-\delta_b\right)\right]\frac{t}{\tau}.
\label{second}
\end{eqnarray}
The calculation was done using a computationally efficient method
described in \ref{app}. The MSD is a monotonically decreasing
function of the density for $\delta_{f}-\delta_{b}>-\frac{1}{3}$,
while for $\delta_{f}-\delta_{b}<-\frac{1}{3}$ it has a single peak at
$\rho=\frac{3}{4}+\left[4\left(\delta_{f}-\delta_{b}\right)\right]^{-1}$.

Not surprisingly, based on a mean field description equations (\ref{first}) and
(\ref{second}) can only give a qualitative picture of the mechanisms behind the
non-monotonicity of the diffusivity. In particular, in 1D the scaling of the MSD
with time is different. Even in 2D, the diffusion coefficient obtained from equation
(\ref{second}) does not agree too well with the simulation results. Furthermore,
while the mean field result accounts for the single maximum scenario and therefore
provides some added value in understanding our observations above, it cannot
capture the strongly non-monotonic regime. Less severely, it does not provide the
correct value of the persistence at the critical point. Moreover, we see from
figure (\ref{fig_deff}b) that the behaviour in 2D depends on $\delta_f$ and
$\delta_b$ in a more complicated manner than simply as $\delta_f-\delta_b$.

\section{Full persistence}
\label{sec_tp}

We now investigate the limiting case of full persistence in 1D. In this
case, the model may be thought of as a two-species totally antisymmetric
exclusion principle (TASEP), with equal populations of right-moving and
left-moving particles. In this model, all motion stops after a short,
density-dependent relaxation time, since a particle stops moving as soon
as it encounters a block containing at least one other particle of the
opposite species.

In order to investigate the motion of, say, a right-moving particle, it
is sufficient to consider its nearest left-moving particle to the right
of it and all the intervening right-moving particles, since all the other
particles cannot affect its motion. Let us consider a right-moving particle
starting from the origin. At time $t=0$, its nearest left-moving particle
(to the right of it) is located at site $n$ with probability
\begin{equation}
q(n)=\left(1-\frac{\rho}{2}\right)^{n-1}\frac{\rho}{2}.
\end{equation}
Let us assume at first that there are no other right-moving particles between
them. Note that until they encounter each other, the movement of the two
particles is completely uncorrelated. Therefore, the probability that the
right-moving particle is at site $m$ at time $t$ given that its nearest
left-moving particle started at $n$, $p(m,n,t)$, is given by
\begin{eqnarray}
\fl p(m,n,t)=&p_{0}(m,t)\sum^{n-m-1}_{m'=0}p_{0}(m',t)\nonumber\\
&+\int^{t}_{0}dt'\int^{t}_{t'}dt''p_{0}(m,t')\left(1-e^{-(t''-t')/\tau}\right)p_{0}(n-m-1,t'')\nonumber\\
&+\int^{t}_{0}dt'\int^{t}_{t'}dt''p_{0}(m,t'')\left(1-e^{-(t''-t')/\tau}\right)p_{0}(n-m-1,t') ,\label{pm}
\end{eqnarray}
where $p_{0}(m,t)$ is the probability that a single independent walker moved
$m$ steps at time $t$. The first term corresponds to the probability that
at time $t$ the right-moving walker reaches site $m$ and the left-moving
walker reached at most site $n-m-1$, so they did not interact yet. The
second term is the probability that the right-moving walker reached site
$m$ at some time $t'$, and did not move until time $t''$ at which point the
left-moving particle reached site $n-m-1$. The last term is analogous to the
second term, with the left-moving particle arriving first. The probability
for an independent walker, $p_{0}(m,t)$, is governed by the evolution equation
\begin{equation}
\tau\frac{\partial p_{0}(m,t)}{\partial t}=-p_{0}(m,t)+p_{0}(m-1,t) ,
\end{equation}
with the initial condition $p_{0}(m,0)=\delta_{m,0}$, and therefore
\begin{equation}
p_{0}(m,t)=e^{-t/\tau}\left(\frac{t}{\tau}\right)^{m}\frac{1}{m!} .\label{p0m}
\end{equation}
Using (\ref{p0m}) in (\ref{pm}) yields
\begin{equation}
p(m,n,t)=\frac{\left(n+1\right)\left[n!-\Gamma\left(n+1,\frac{2t}{\tau}\right)\right]}{2^{n+1}\left(n-m\right)!\left(m+1\right)!}+e^{-t/\tau}\left(\frac{t}{\tau}\right)^{m}\frac{\Gamma\left(n-m,\frac{t}{\tau}\right)}{m!\left(n-m-1\right)!} ,
\end{equation}
where $\Gamma(n,z)$ is the incomplete gamma function \cite{IncompleteGamma}. Summing over all possible initial locations for the left moving particle yields
\begin{eqnarray}
\fl p(m,t)=\sum^{\infty}_{n=m+1}p(m,n,t)q(n)=\rho\left(1-\frac{\rho}{2}\right)^{m-1}\left(\frac{1}{\left(1+\frac{\rho}{2}\right)^{m+2}}-\frac{1}{2^{m+2}}\right)\nonumber\\
\fl+e^{-\left(1+\rho/2\right)t/\tau}\left(\frac{t}{\tau}\right)^{m}\left(1-\frac{\rho}{2}\right)^{m}\frac{1}{(m+1)!}\left(m+1-\frac{\rho}{2+\rho}\frac{t}{\tau}\right)\nonumber\\
\fl+\frac{\rho}{m!\left(4-\rho^{2}\right)}\left(\frac{t}{\tau}\right)^{m+1}\left(1-\frac{\rho}{2}\right)^{m}\left[\left(2+\rho\right)E_{-m}\left(\frac{2t}{\tau}\right)-4E_{-m}\left(\frac{t(1+\rho/2)}{\tau}\right)\right] ,
\end{eqnarray}
where $E_{\nu}(z)$ is the exponential integral $E$ \cite{ExponentialIntegral}
\begin{equation}
E_{\nu}(z)=\int^{\infty}_{1}\frac{e^{-zt}}{t^{\nu}}dt .
\end{equation}

\begin{figure}
\centering
\includegraphics[height=4.8cm]{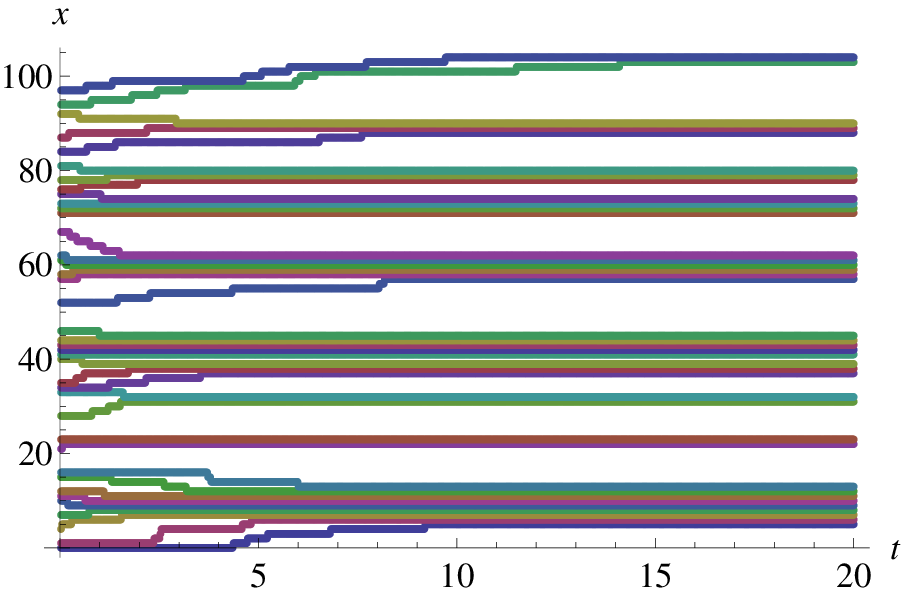}
\includegraphics[height=4.8cm]{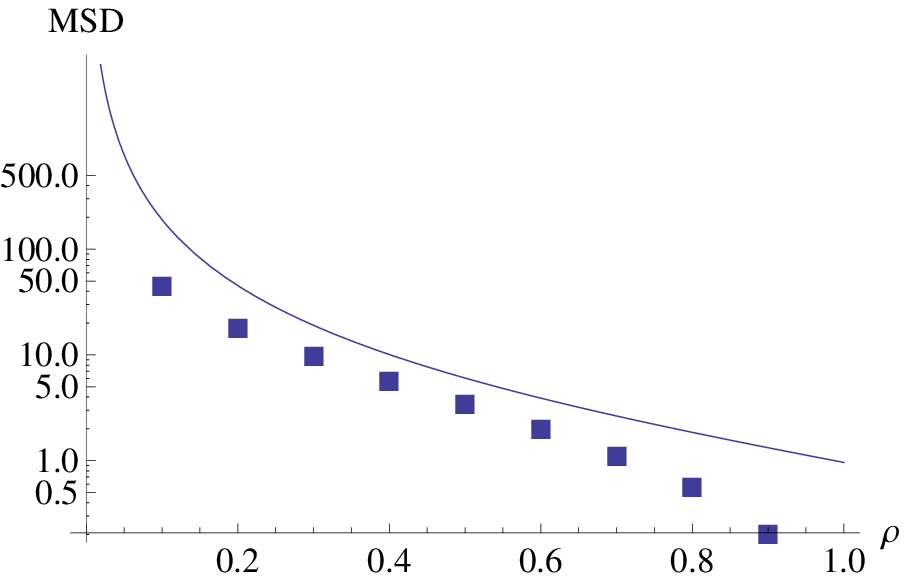}
\caption{Totally persistent case. Left: sample trajectory of the system.
Right: Long time value
of the MSD versus density. The symbols are simulations results, the continuous
line is the analytical upper bound (\ref{msdtp}).}
\label{fig_msd_tp}
\end{figure}

The MSD of such a right-moving particle without any intervening other right-moving particles is thus
\begin{eqnarray}
\left\langle x^{2}\right\rangle_{0}=\sum^{\infty}_{n=1}m^{2}p(m,t)=\frac{2\left(2-\rho\right)\left(4+6\rho+4\rho^{2}-\rho^{3}\right)}{\rho^{2}\left(2+\rho\right)^{3}}\nonumber\\
+\left[\frac{2\rho}{4-\rho^{2}}+\frac{32\rho\left[2-3\rho-2\left(2-\rho\right)\left(1-\rho\right)\frac{t}{\tau}\right]}{\left(2-\rho\right)^{3}\left(2+\rho\right)\left[2-\left(2-\rho\right)\frac{t}{\tau}\right]^{2}}\right]e^{-(1+\rho/2)t/\tau}\nonumber\\
+\frac{e^{-\rho t/\tau}}{4\left(-4+\rho^{2}\right)}\left\{8\rho-2\left[8-\left(4-\rho\right)\rho^{2}\right]\frac{t}{\tau}-\left(2-\rho\right)^{2}\left[4-\rho\left(2+\rho\right)\right]\left(\frac{t}{\tau}\right)^{2}\right\}\nonumber\\
+\frac{4\rho\left[4\rho+\left(2-\rho\right)\left(2-3\rho\right)\frac{t}{\tau}\right]}{\left(2-\rho\right)^{3}\left[2-\left(2-\rho\right)\frac{t}{\tau}\right]^{2}}e^{-2t/\tau}\nonumber\\
+\frac{2e^{-4/(2-\rho)}\rho}{\left(2-\rho\right)^{4}}\left[4-\rho\left(8+\rho\right)\right]\mathrm{Ei}\left(\frac{4}{2-\rho}-\frac{2t}{\tau}\right)\nonumber\\
+\frac{8e^{-(2+\rho)/(2-\rho)}\rho\left(-4+5\rho^{2}\right)}{\left(2-\rho\right)^{4}\left(2+\rho\right)}\mathrm{Ei}\left[\frac{2+\rho}{2-\rho}-\left(1+\frac{\rho}{2}\right)\frac{t}{\tau}\right] ,\label{msdtp}
\end{eqnarray}
where $\mathrm{Ei}(z)$ is the exponential integral $\mathrm{Ei}$ \cite{ExponentialIntegral}
\begin{equation}
\mathrm{Ei}(z)={\cal P}\int^{z}_{-\infty}\frac{e^{t}}{t}dt .
\end{equation}
In the long time limit the MSD converges to a constant, given by the first
line of (\ref{msdtp}).  If there are intervening particles, then the MSD must
be lower than that given by (\ref{msdtp}). Therefore, the MSD of a totally
persistent system converges to a density-dependent constant, bounded from
above by (\ref{msdtp}). Figure \ref{fig_msd_tp} shows the value of the MSD
at long times compared to the upper bound. We observe that the bound is
indeed fulfilled. Remarkably, the upper bound provides a fairly good
approximation to the simulated data at intermediate densities.

\section{Full anti-persistence}
\label{sec_tap}

\begin{figure}
\centering
\includegraphics[height=4.4cm]{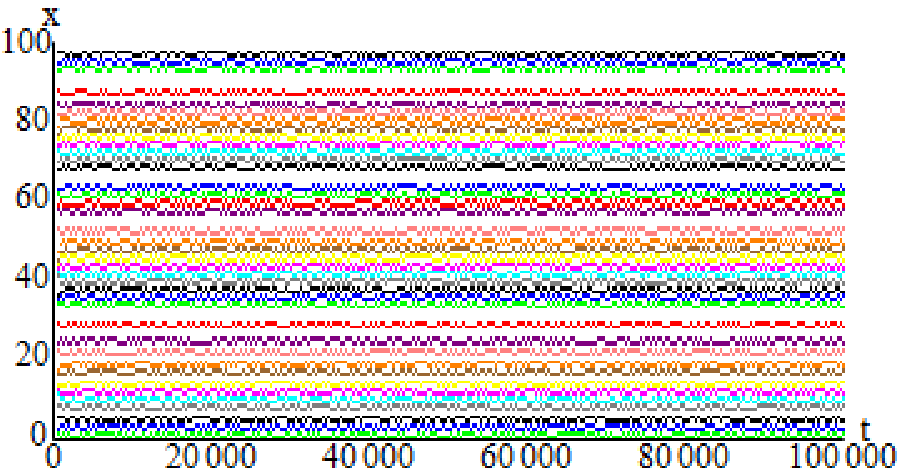}
\includegraphics[height=4.4cm]{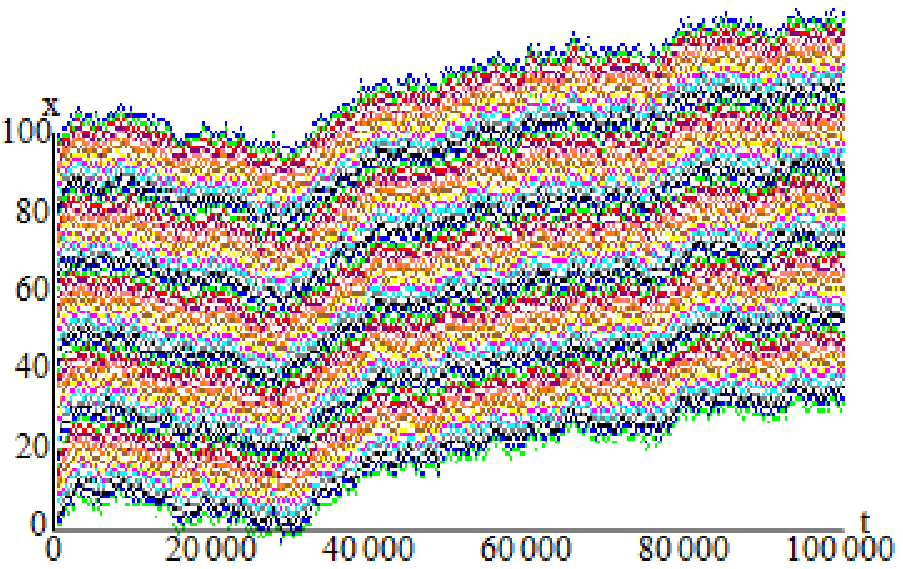}
\caption{Time evolution of a fully anti-persistent system of length $100$ with
periodic boundary conditions, with either density $\rho=0.4$ (left panel) or
$\rho=0.6$ (right panel). Each colour represents a different particle, and
vacancies are represented in white. At low density, the particles are localised,
while at the high density they move.}
\label{tap_conf}
\end{figure}

In the complementary limit of full anti-persistence the system exhibits
several unique properties. We consider a 1D lattice with totally
anti-persistent particles, such that at each step the particles always switch
direction and attempt to move in the opposite direction than before. To
our knowledge, this pathological case has not been explored before. The
two sites between which the particle hops change only if another particle
enters one of the two sites, such that the first walker pushes itself on
its new neighbour. Physically, this limit represents very deep and narrow
traps, such that a particle can escape only if another particle enters its
trap. In a closed system with density $\rho<1/2$, we find that the system
relaxes to a steady state in which each particle jiggles between two sites,
and thus the MSD converges to a constant, see figure \ref{tap_conf}.

\begin{figure}
\includegraphics[height=3.6cm]{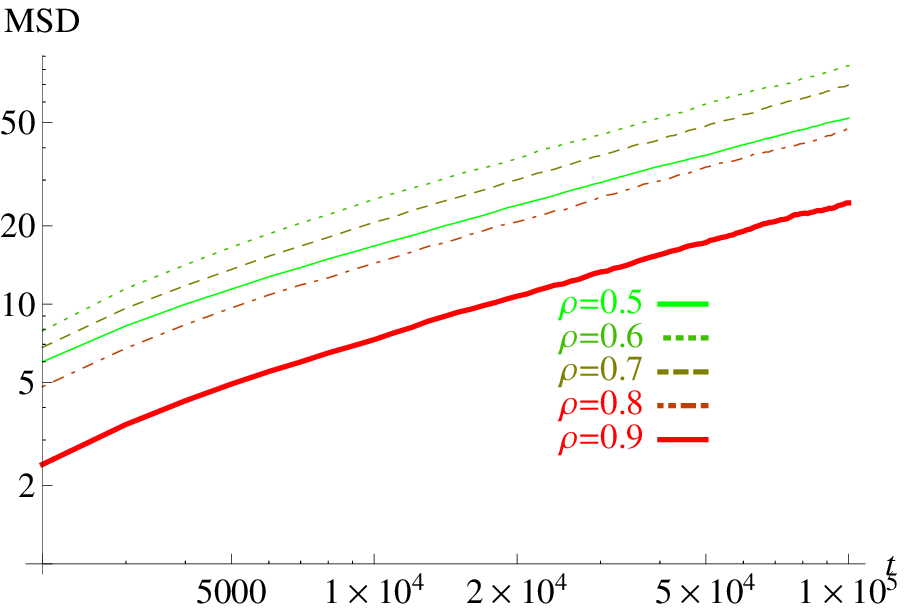}
\includegraphics[height=3.6cm]{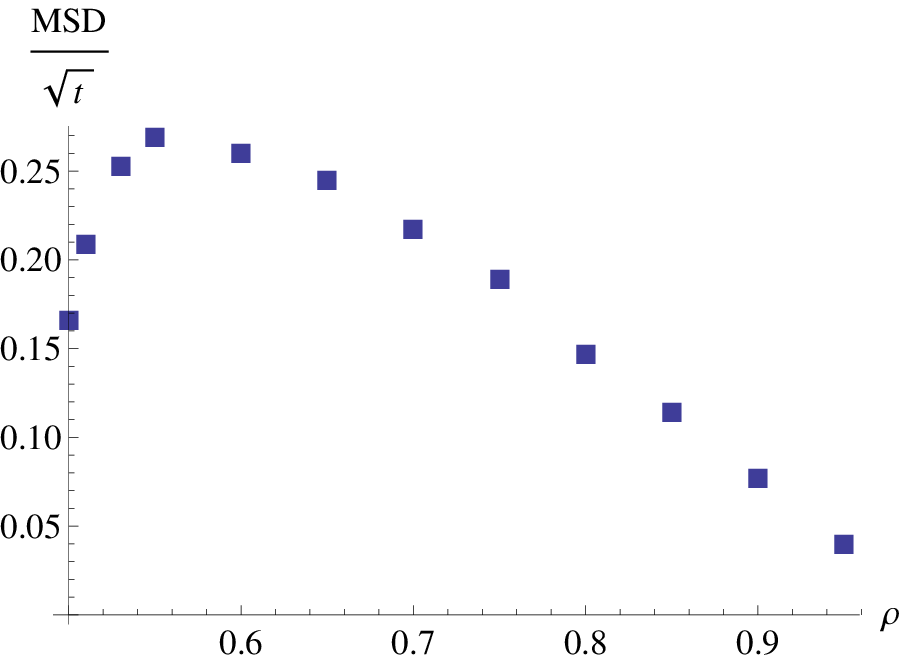}
\includegraphics[height=3.6cm]{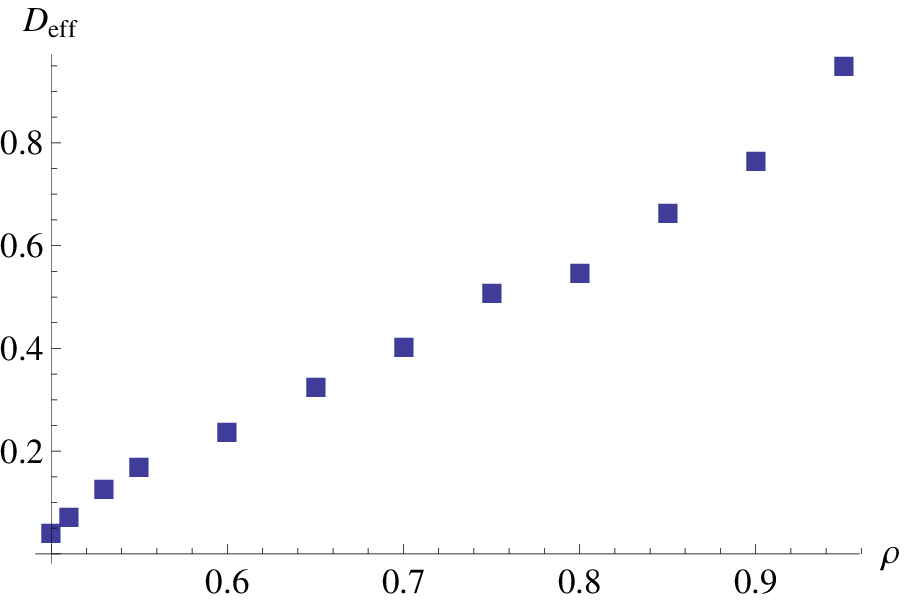}
\caption{MSD and effective diffusion coefficient in the fully anti-persistent
case for density $\rho\geq0.5$.}
\label{tap_msd}
\end{figure}

For $\rho>1/2$ we find that the MSD scales as $\sqrt{t}$, as shown in figure
\ref{tap_msd}a. We extract the effective diffusion coefficient
$D_{\mathrm{eff}}$ from the slope of the MSD vs. time and show the results in figure
\ref{tap_msd}b-c. We note that unlike in the case with
$\delta>-1/2$, here the MSD divided by $\sqrt{t}$ has a single maximum as a
function of the density. At a density of exactly $\rho=1/2$, the MSD appears
to grow slightly slower than $\sqrt{t}$, however more precise measurements
are required to find its exact time dependence.

\section{Conclusions}
\label{sec_summary}

We studied a lattice gas of persistent walkers in which each site is occupied
by at most one particle and the direction each particle attempts to move to
depends on its last step. The directionality is modulated by a "persistence",
the particle's tendency to continue moving in the same direction as in the
last step.  Specifically, we analysed the mean squared displacement (MSD)
as function of the particle density and the persistence.

First, we found that the scaling of the MSD with time is the same as in
memory-less systems, i.e., the simple symmetric exclusion principle (SSEP)
model, MSD$\propto\sqrt{t}$ in 1D and MSD$\propto t$ in 2D. We expect that
this scaling remains true even for particles with longer memory, as long as
the velocity autocorrelations decay sufficiently fast.

Second, we observed that the MSD increases with growing persistence of the
walkers, in accordance with intuition. However, the dependence of the MSD on
the density turns out to be non-trivial. As long as the anti-persistence,
i.e. the tendency to go backwards, is not too high, the MSD decreases
with the density $\rho$ simply due to the crowdedness. However, for highly
anti-persistent walkers, the MSD \emph{increases\/} with growing $\rho$ for
low densities and reaches a \emph{maximum\/} at some persistence-dependent
density. This occurs because for highly anti-persistent walkers, the other
walkers prevent a given walker from stepping backwards and thus effectively
increase their MSD. There is also an intermediate regime in which for low
densities the MSD decreases with the density, but at intermediate densities
it reaches both a minimum and a maximum. To our knowledge, this type of
behaviour has not been observed before.

Third, we considered the two extreme limits of full persistence and
anti-persistence. For totally persistent particles, although a single walker
performs a ballistic motion, when other particles come into play, the single
walker is blocked by other particles going in the opposite direction, such
that the MSD saturates and all movement halts. We derived an upper bound for
the MSD in this case which quite nicely approximates the simulations results
at intermediate particle densities. In the totally anti-persistent case,
a single particle jumps between two sites, but when other particles exist
and the density is higher than $\frac{1}{2}$, they enable it to move and
the MSD grows as $\sqrt{t}$. We suspect that at exactly $\rho=\frac{1}{2}$
the MSD grows with time slower than $\sqrt{t}$, but more precise measurements
are needed.

An interesting expansion of this work would be to analyse the MSD of walkers
with slowly decaying velocity autocorrelations, in which the MSD of a
single walker is not linear in time (anomalous diffusion \cite{pccp}). In
1D a similar model was investigated in \cite{Sanders2014}. We expect that
for positive autocorrelations (i.e. persistent walkers) the MSD in 1D
scales as $\sqrt{t}$ for all densities, while for negative autocorrelations
(i.e. anti-persistent walkers) it scales as $\sqrt{t}$ for $\rho>\frac{1}{2}$
and slower for $\rho<\frac{1}{2}$.
From preliminary results on crowded L{\'e}vy walkers, we indeed find that in 1D
the MSD scales as $\sqrt{t}$ and in 2D it scales as $t$. Furthermore, at high
enough densities the system exhibits motiliy induced phase seperation (MIPS).

In the model we investigated here, the direction chosen at each step is
completely uncorrelated to the success of failure of the moves and thus to
the other particles. Another interesting expansion involves correlating the
chosen direction with the density, such that the probability to move forward or
backwards depends on whether the last move was successful or not. We conjecture
that persistent walkers that tend to turn around when they are blocked would
exhibit a transition from a behaviour similar to positive persistence at low
densities to a behaviour similar to anti-persistence at high densities.
Another expansion of this
model would be to add quenched disorder to the system, i.e., to fix some of the
particles in place so they become obstacles. It is possible that at sufficiently
high densities, perhaps related to the percolation threshold, anti-persistent
particles will have a higher MSD than persistent particles.

\ack

We thank Yair Shokef, Michael Urbakh and Andrey Cherstvy for fruitful
discussions. ET acknowledges financial support from the TAU-Potsdam
fellowship. RM acknowledges the Foundation for Polish Science for funding
within an Alexander von Humboldt Polish Research Scholarship, and DFG within
grants ME 1535/6-1 and 1535/7-1.

\appendix

\section{Calculation of the MSD in a mean field approximation}
\label{app}

We here present a computationally efficient method of calculating
the MSD in the mean field approximation. We derive the result for
1D systems, but expanding it to higher dimensions is straightforward.
The derivation applies to more general models than we consider in this
work.

Consider a single particle with a set of internal states moving in
1D. It is not constrained to move on a lattice. Given that it is
currently at state $\eta$, it moves a distance $r$ and changes its state to
$\eta'$ with rate ${\cal M}_{\eta,\eta'}(r)/\tau$. The evolution equation
may be written in matrix form as
\begin{equation}
\tau\frac{\partial\textbf{Q}(x,t)}{\partial t}=\int^{\infty}_{-\infty}dr{\cal
M}(r)\textbf{Q}\left(x-r,t\right) .
\end{equation}
In Fourier space the evolution equation is given by
\begin{equation}
\tau\frac{\partial\tilde{\textbf{Q}}(k,t)}{\partial t}={\cal M}(k)\tilde{
\textbf{Q}}(k,t) ,\label{fourq}
\end{equation}
When $k=0$, the dynamical matrix has a single zero eigenvalue, with
the corresponding right and left eigenvectors $\textbf{V}_{0}$ and
$\textbf{U}^{T}_{0}$. Note that all entries of the left eigenvector are
equal to $1$, and the eigenvector $\textbf{V}_{0}$ represents the steady
state of the system. The other eigenvalues and eigenvectors are denoted by
$\lambda_{i}$, $\textbf{V}_{i}$ and $\textbf{U}^{T}_{i}$ with $i>0$.
The real part of all the other eigenvalues is negative. The solution to
equation (\ref{fourq}) is
\begin{equation}
\tilde{\textbf{Q}}(k,t)=e^{{\cal M}(k)t}\tilde{\textbf{Q}}(k,0) .
\end{equation}
We assume that at time $t=0$ the system is in
the steady state and the particle is at the origin
(i.e. $\textbf{Q}(n,0)=\delta_{n,0}\textbf{V}_{0}$). Therefore, we note that
at time $t=0$ the vector $\tilde{\textbf{Q}}(k,0)$ is independent of $k$,
since by definition
\begin{equation}
\tilde{\textbf{Q}}(k,0)=\sum^{\infty}_{-\infty}e^{ink}\textbf{Q}(n,0)=\sum^{
\infty}_{-\infty}e^{ink}\delta_{n,0}\textbf{V}_0=\textbf{V}_0 .
\end{equation}
The MSD is given by
\begin{equation}
\left\langle n^{2}\right\rangle=-\left.\frac{\partial^{2}}{\partial k^{2}}\sum_{
\sigma,\sigma_{-},\sigma_{+}}\tilde{Q}_{\sigma,\sigma_{-},\sigma_{+}}(k,t)\right|_{
k=0}=-\left.\frac{\partial^{2}}{\partial k^{2}}\textbf{U}^{T}_{0}e^{{\cal M}(k)t/
\tau}\textbf{V}_{0}\right|_{k=0} .
\end{equation}
We now use \cite{Wilcox1967}
\begin{equation}
\frac{\partial}{\partial k}e^{{\cal M}(k)t/\tau}=\frac{t}{\tau}\int^1_0e^{\alpha{
\cal M}(k)t/\tau}\frac{\partial{\cal M}(k)}{\partial k}e^{\left(1-\alpha\right){
\cal M}(k)t/\tau}d\alpha,
\end{equation}
and find that
\begin{eqnarray}
\langle n^2\rangle&=&-\frac{t}{\tau}\textbf{U}^{T}_{0}\int^{1}_{0}\left\{\left[
\int^{1}_{0}\alpha\frac{t}{\tau}e^{\alpha\beta{\cal M}t/\tau}\frac{\partial{\cal
M}}{\partial k}e^{\alpha\left(1-\beta\right){\cal M}t/\tau}d\beta\right]\frac{
\partial{\cal M}}{\partial k}e^{\left(1-\alpha\right){\cal M}t/\tau}\right.
\nonumber\\
&&\left.+e^{\alpha{\cal M}t/\tau}\frac{\partial{\cal M}}{\partial k}\left[\int^1_0
\left(1-\alpha\right)\frac{t}{\tau}e^{\left(1-\alpha\right)\beta{\cal M}t/\tau}
\frac{\partial{\cal M}}{\partial k}e^{\left(1-\alpha\right)\left(1-\beta\right){
\cal M}t/\tau}d\beta\right]\right.\nonumber\\
&&\left.+e^{\alpha{\cal M}t\tau}\frac{\partial^2{\cal M}}{\partial k^2}e^{\left(
1-\alpha\right){\cal M}t/\tau}\right\}d\alpha\textbf{V}_0,
\end{eqnarray}
where the matrix ${\cal M}$ and its derivatives are evaluated at $k=0$. Since
$\textbf{V}_{0}$ and $\textbf{U}^{T}_{0}$ are the zero eigenvalues of ${\cal M}$
at $k=0$ we find that
\begin{eqnarray}
\langle n^2\rangle&=&-\frac{t}{\tau}\textbf{U}^T_0\int^1_0\left\{\alpha\frac{t}{
\tau}\frac{\partial{\cal M}}{\partial k}\left[\int^1_0e^{\alpha\left(1-\beta\right){
\cal M}t/\tau}d\beta\right]\frac{\partial{\cal M}}{\partial k}\right.\nonumber\\
&&\left.+\frac{\partial{\cal M}}{\partial k}\left(1-\alpha\right)\frac{t}{\tau}
\left[\int^1_0e^{\left(1-\alpha\right)\beta{\cal M}t/\tau}d\beta\right]\frac{
\partial{\cal M}}{\partial k}+\frac{\partial^2{\cal M}}{\partial k^2}\right\}d
\alpha\textbf{V}_0.
\end{eqnarray}
In the first term we change the integration variable from $\beta$ to $1-\beta$,
in the second term we change the integration variable from $\alpha$ to $1-\alpha$,
and in the third term we perform the integration over $\alpha$, such that
\begin{eqnarray}
\left\langle n^{2}\right\rangle=-\frac{t}{\tau}\textbf{U}^{T}_{0}\left\{2\frac{t}{
\tau}\frac{\partial{\cal M}}{\partial k}\left[\int^1_0\int^1_0\alpha e^{\alpha\beta{
\cal M}t/\tau}d\alpha d\beta\right]\frac{\partial{\cal M}}{\partial k}+\frac{
\partial^2{\cal M}}{\partial k^2}\right\}\textbf{V}_0 .
\end{eqnarray}
We now use the spectral decomposition of ${\cal M}$
\begin{equation}
{\cal M}=\sum_{i>0}\lambda_i\textbf{V}_{i}\textbf{U}^T_i ,
\end{equation}
such that
\begin{equation}
e^{\alpha\beta{\cal M}t/\tau}=\textbf{V}_0\textbf{U}^T_0+\sum_{i>0}e^{\alpha
\beta\lambda_{i}t/\tau}\textbf{V}_i\textbf{U}^T_i .
\end{equation}
We may now perform the integrals over $\alpha$ and $\beta$
\begin{equation}
\int^1_0\int^1_0\alpha e^{\alpha\beta{\cal M}t/\tau}d\alpha d\beta=\frac{1}{2}
\textbf{V}_0\textbf{U}^T_0+\sum_{i>0}\textbf{V}_i\textbf{U}^T_i\left[-\frac{1}{
\lambda_it/\tau}+\frac{e^{\lambda_it/\tau}-1}{\left(\lambda_it/\tau\right)^2}\right],
\end{equation}
which in the long time limit is
\begin{equation}
\int^1_0\int^1_0\alpha e^{\alpha\beta{\cal M}t/\tau}d\alpha d\beta=\frac{1}{2}
\textbf{V}_0\textbf{U}^T_0-\sum_{i>0}\textbf{V}_i\textbf{U}^T_i\frac{1}{\lambda_i
t/\tau}.
\end{equation}
We define a new matrix
\begin{equation}
\tilde{{\cal M}}={\cal M}+\textbf{V}_0\textbf{U}^T_0,
\end{equation}
such that
\begin{eqnarray}
\fl\int^1_0\int^1_0\alpha e^{\alpha\beta{\cal M}t/\tau}d\alpha d\beta&=\frac{
1}{2}\textbf{V}_0\textbf{U}^T_0+\textbf{V}_0\textbf{U}^T_0\frac{1}{t/\tau}-
\sum_{i>0}\textbf{V}_i\textbf{U}^T_i\frac{1}{\lambda_it/\tau}-\textbf{V}_0
\textbf{U}^T_0\frac{1}{t/\tau}\nonumber\\
\fl&=\left(\frac{1}{2}+\frac{\tau}{t}\right)\textbf{V}_0\textbf{U}^T_0-\frac{
\tau}{t}\tilde{{\cal M}}^{-1} ,
\end{eqnarray}
and thus
\begin{eqnarray}
\left\langle n^{2}\right\rangle=&-2\left(\frac{1}{2}+\frac{\tau}{t}\right)\left(\frac{t}{\tau}\right)^{2}\left(\textbf{U}^{T}_{0}\frac{\partial{\cal M}}{\partial k}\textbf{V}_{0}\right)^{2}\nonumber\\
&+2\frac{t}{\tau}\textbf{U}^{T}_{0}\frac{\partial{\cal M}}{\partial k}\tilde{{\cal M}}^{-1}\frac{\partial{\cal M}}{\partial k}\textbf{V}_{0}-\frac{t}{\tau}\textbf{U}^{T}_{0}\frac{\partial^{2}{\cal M}}{\partial k^{2}}\textbf{V}_{0} .
\end{eqnarray}
Due to symmetry the first term vanishes and thus
\begin{equation}
\left\langle n^{2}\right\rangle=2\frac{t}{\tau}\textbf{U}^{T}_{0}\frac{\partial{\cal M}}{\partial k}\tilde{{\cal M}}^{-1}\frac{\partial{\cal M}}{\partial k}\textbf{V}_{0}-\frac{t}{\tau}\textbf{U}^{T}_{0}\frac{\partial^{2}{\cal M}}{\partial k^{2}}\textbf{V}_{0} .\label{msd_app}
\end{equation}
Note that in the case of isotropic one-step memory, all elements of the steady
state distribution $\textbf{V}_0$ are equal, and thus constructing the matrix
$\textbf{V}_0\textbf{U}^T_0$ is trivial, and the only time consuming part
of the calculation is the inversion of the matrix $\tilde{{\cal M}}$. It
is straightforward to check that the MSD in a $d$ dimensional system is $d$
times the expression in (\ref{msd_app}).


\section*{References}

\begin{thebibliography}{99}

\bibitem{ref1}{de Groot B L and Grubm{\"u}ller 2005 \textit{Curr. Opin. Struct.
Biol.} \textbf{15} 176.\\
H{\"o}fling F and Franosch T 2014 \textit{Rep. Prog. Phys.} \textbf{76} 046602.\\
Bechinger C, Di Leonardo R, L{\"o}wen H, Reichhardt C, Volpe G and Volpe G 2016
\textit{Rev. Mod. Phys.} \textbf{88} 045006.\\
Metzler R, Jeon J H and Cherstvy A G 2016 \textit{BBA-Rev. Biomembranes}
\textbf{1858} 2451.\\
Hakim V and Silberzan P 2017 \textit{Rep. Prog. Phys.} \textbf{80} 076601.\\
N{\o}rregaard K, Metzler R, Ritter C M, Berg-S{\o}rensen K and Oddershede L B 2017
\textit{Chem. Rev.} \textbf{117} 4342}.

\bibitem{Hasnain2015}{Hasnain S and Bandyopadhyay P 2015 \textit{J. Chem. Phys.}
\textbf{143} 114104}.

\bibitem{ref2}{Detcheverry F 2015 \textit{Europhys. Lett.} \textbf{111} 60002.\\
Rupprecht J, B{\'e}nichou O and Voituriez R 2016 \textit{Phys. Rev. E} \textbf{94}
012117.\\
Detcheverry F 2017 \textit{Phys. Rev. E} \textbf{96} 012415}.

\bibitem{ref3}{Sevilla F 2016 \textit{Phys. Rev. E} \textbf{94} 062120.\\
Ariel G, Be'er A and Reynolds A 2017 \textit{Phys. Rev. Lett.} \textbf{118} 228102.\\
Fedotov S and Korabel N 2017 \textit{Phys. Rev. E} \textbf{95} 030107}.

\bibitem{ref4}{Wioland H, Lushi E and Goldstein R 2016 \textit{New J Phys.}
\textbf{18} 075002.\\
Stenhammer J, Nardini C, Nash R, Marenduzzo D and Mozorov A 2017 \textit{Phys.
Rev. Lett.} \textbf{119} 028005}.

\bibitem{Berthier2013}{Berthier L and Kurchan J 2013 \textit{Nat. Phys.}
\textbf{9} 310}.

\bibitem{ref5}{Viscek T, Czir{\'o}k A, Ben-Jacob E, Cohen I and Shochet O
1995 \textit{Phys. Rev. Lett.} \textbf{75} 1226.\\
Sep{\'u}lveda N, Petitjean L, Cochet O, Grasland-Mongrain E, Silberzan P
and Hakim V 2013 \textit{PLoS Comput. Biol.} \textbf{9} 1002944.\\
Gro{\ss}mann R, Peruani F and B{\"a}r M 2016 \textit{Phys. Rev. E} \textbf{94}
050602.\\
Liebchen B and Levis D 2017 \textit{Phys. Rev. Lett.} \textbf{119} 058002}.

\bibitem{Zimmermann2016}{Zimmerman J, Camley B, Rappel W and Levine H 2016
\textit{Proc. Natl. Acad. Sci. USA} \textbf{113} 2660}.

\bibitem{ref6}{Farhadifar R, R{\"o}per J, Aigouy B, Eaton S and J{\"u}licher F
2007 \textit{Curr. Biol.} \textbf{17} 2095.\\
Staple D, Farhadifar R, R{\"o}per J, Aigouy B, Eaton S and J{\"u}licher F 2010
\textit{Eur. Phys. J E} \textbf{33} 117.\\
S{\'a}ndor C, Lib{\'a}l A, Reichhardt C ad Reichhardt C 2017 \textit{Phys. Rev. E}
\textbf{95} 032606}.

\bibitem{Reichhardt2014}{Reichhardt C and Reichhardt C 2014 \textit{Soft Matter}
\textbf{10} 7502}.

\bibitem{Zachreson2017}{Zacherson C, Wolff C, Whitchurch C and Toth M 2017
\textit{Phys. Rev. E} \textbf{95} 012408}.

\bibitem{Reichhardt2014b}{Reichhardt C and Reichhardt C 2014 \textit{Phys. Rev. E}
\textbf{90} 012701}.

\bibitem{Lam2015}{Lam K, Schindler M and Dauchot O 2015 \textit{New J Phys.}
\textbf{17} 113056}.

\bibitem{Graf2017}{Graf I and Frey E 2017 \textit{Phys. Rev. Lett.} \textbf{188}
128101}.

\bibitem{Illien2015}{Illien P, B{\'e}nichou O, Oshanin G and Voituriez R 2015
\textit{J Stat. Mech.} P11016}.

\bibitem{ref7}{Mark S, Shlomovitz R, Gov N, Poujade M, Grasland-Mongrain E and
Silberzan P 2010 \textit{Biophys. J} \textbf{98} 361.\\
McCalla S and Brecht J 2016 \textit{Phys. Rev. E} \textbf{94} 060401}.

\bibitem{Fisher2014}{Fisher H, Giomi L, Hoekstra H and Mahadevan L 2014
\textit{Proc. Roy. Soc. B} \textbf{281} 20140296}.

\bibitem{Berg1990}{Berg H C and Turner L 1990 \textit{Biophys. J.} \textbf{58} 919}.

\bibitem{Schulz2011}{Schulz J, Kolomeisky A B and Frey E 2011 \textit{Europhys.
Lett.} \textbf{95} 30004}.

\bibitem{Ghosh2015b}{Ghosh S K, Cherstvy A G and Metzler R 2015 \textit{Phys.
Chem. Chem. Phys.} \textbf{17} 1847}.

\bibitem{Hermann2017}{Hermann C J J, Metzler R and Engbert R 2017 \textit{Sci.
Rep.} \textbf{7} 12958}.

\bibitem{Taylor1921}{Taylor G 1921 \textit{P Lond. Math. Soc.} \textbf{2} 196}.

\bibitem{Tchen1952}{Tchen C 1952 \textit{J Chem. Phys.} \textbf{20} 214}.

\bibitem{Kareiva1983}{Kareiva P and Shigesada N 1983 \textit{Oecologia} \textbf{56}
234}.

\bibitem{Boguna1998}{Bogu{\~n}a M, Porr{\'a} J and Masoliver J 1998 \textit{Phys.
Rev. E} \textbf{58} 6992}.

\bibitem{Romanczuk2012}{Romanczuk P, B{\"a}r M, Ebeling W, Lindner B and
Schminasky-Geier L 2012 \textit{Eur. Phys. J} \textbf{202} 1}.

\bibitem{Ghosh2015}{Ghosh P, Li Y, Marchgiani G and Marchesoni F 2015 \textit{J.
 Chem. Phys.} \textbf{143} 211101}.

\bibitem{Tahir}{Tahir-Kheli R A and Elliot R J 1983 \textit{Phys. Rev. B}
\textbf{27} 844; Tahir-Kheli R A 1983 \textit{Phys. Rev. B} \textbf{27} 7229}.

\bibitem{Spitzer1970}{Spitzer F 1970 \textit{Adv. Math.} \textbf{5} 246}.

\bibitem{Melbinger2011}{Melbinger A, Reichenbach T, Franosch T and Frey E 2011
\textit{Phys. Rev. E} \textbf{83} 031923}.

\bibitem{vdz} Gorissen M, Lazarescu A, Mallick K, and C. Vanderzande C 2012
\textit{Phys. Rev. Lett.} \textbf{109}, 170601.

\bibitem{ref8}{Richards P M 1977 \textit{Phys. Rev. B} \textbf{16} 1393.\\
Brak R and Elliot R J 1989 \textit{J Phys. Condens. Matter} \textbf{1} 10299.\\
Derrida M, Evans M R, Hakim V and Pasquier V 1993 \textit{J Phys. A} \textbf{26}
1493.\\
Honecker A and Peschel I 1997 \textit{J Stat. Phys.} \textbf{88} 319.\\
Lazarescu A and Mallick K 2011 \textit{J Phys. A} \textbf{44} 315001}.

\bibitem{Waghe2012}{Waghe A, Rasaiah J C and Hummer G 2012 \textit{J. Chem.
Phys.} \textbf{137} 044709}.

\bibitem{Yang2010}{Yang S Y, Yang J A, Kim E S, Jeon G, Oh E J, Choi K Y, Hahn
S K and Kim J K 2010 \textit{ACS Nano} \textbf{4} 3817}.

\bibitem{jae} Song MS, Moon HC, Jeon JH, and Park  HY 2018, \textit{Nature Comm.}
\textbf{9}, 344.\\
Chen KJ, Wang B, and Granick S 2015, \textit{Nature Mater.} \textbf{14}, 589.

\bibitem{Harris1965}{Harris T 1965 \textit{J Appl. Probab.} \textbf{2} 323}.

\bibitem{Spohn1983}{Spohn H 1983 \textit{J Phys. A} \textbf{16} 4275}.

\bibitem{ref10}{Illien P, B{\'e}nichou O, Mej{\`i}a-Monasterio C, Oshanin G
and Voituriez R 2013 \textit{Phys. Rev. Lett.} \textbf{111} 038102.\\
B{\'e}nichou O, Illien P, Oshanin G, Sarracino A and Voituriez R 2014
\textit{Phys. Rev. Lett.} \textbf{1113} 268002.\\
Illien P, B{\'e}nichou O, Oshanin G and Voituriez R 2015 \textit{J Stat. Mech.}
\textbf{P11016}.\\
B{\'e}nichou O, Illien P, Oshanin G, Sarracino A and Voituriez R 2018
\textit{J Phys. Condens. Matter} \textbf{30} 443001}.

\bibitem{Markham2013}{Markham D C, Simpson M J, Maini P K, Gaffney E A and
Baker R E 2013 \textit{Phys. Rev. E} \textbf{88} 052713}.

\bibitem{Arita2014}{Arita C, Krapivsky P L and Mallick K 2014 \textit{Phys.
Rev. E} \textbf{90} 052108}.

\bibitem{Szavitz2018}{Szavitz-Nossan J, Romano M C and Ciandrini L 2018
\textit{Phys. Rev. E} \textbf{97} 052139}.

\bibitem{Ritort2003}{Ritort F and Sollich P 2003 \textit{Adv. in Phys.}
\textbf{52} 219}.

\bibitem{Whitelam2017}{Whitelam S, Klymko K and Mandal D 2017 \textit{J Chem.
Phys.} \textbf{148} 154902}.

\bibitem{gol} Soto R and Golestanian R, \textit{Phys. Rev. E} \textbf{89},

\bibitem{Treloar2011}{Treloar K K, Simpson M J and McCue S W 2011 \textit{Phys.
Rev. E} \textbf{84} 061920}.

\bibitem{Kourbane2018}{Kourbane-Houssene M, Erignoux C, Bodineau T and Tailleur
J 2018 \textit{Phys. Rev. Lett.} \textbf{120} 268003}.

\bibitem{Manacorda2017}{Manacorda A and Puglisi A 2017 \textit{Phys. Rev. Lett.}
\textbf{119} 208003}.

\bibitem{Gavagnin2018}{Gavagnin E and Yates C A 2018 \textit{Phys. Rev. E}
\textbf{97} 32416}.

\bibitem{Arita2018}{Arita C and Ragoucy E 2018 \textit{Phys. Rev. E} \textbf{98}
052118}.

\bibitem{Galanti2013}{Galanti M, Fanelli D and Piazza F 2013 \textit{Eur. Phys.
J. B} \textbf{86} 456}.

\bibitem{Sanders2014}{Sanders L, Lomholt M, Lizana L, Fogelmark K, Metzler R
and Ambj{\"o}rnsson 2014 \textit{New J Phys.} \textbf{16} 113050}.

\bibitem{Bertrand2018}{Bertrand T, Illien P, B{\'e}nichou O and Voituriez R 2018
\textit{New J Phys.} \textbf{20} 113045}.

\bibitem{Chatterjee2018}{Chatterjee R, Segall N, Merrigan C, Ramola K, Chakraborty
B, and Shokef Y 2018 ArXiv:1812.0614}.

\bibitem{ref9}{Nakazato K and Kitahara K 1980 \textit{Prog. Theor. Phys.}
\textbf{64} 2261.\\
Landim C, Olla S and Varadhan S R S 2001 \textit{Commun. Math. Phys.}
\textbf{224} 307}.

\bibitem{Illien2018}{Illien P, B{\'e}nichou O, Oshanin G, Sarracino A and Voituriez
R 2018 \textit{Phys. Rev. Lett.} \textbf{120} 200606}.

\bibitem{Lizana2010}{Lizana L, Ambj{\"o}rnsson T, Taloni A, Barkai E and Lomholt
M A 2010 \textit{Phys. Rev. E} \textbf{81} 051118}.

\bibitem{IncompleteGamma}{Abramowitz M and Stegun I A 1972, \textit{Handbook of
Mathematical Functions} (Washington DC: National Bureau of Standards), Chapter 6.5}.

\bibitem{ExponentialIntegral}{Abramowitz M and Stegun I A 1972, \textit{Handbook
of Mathematical Functions} (Washington DC: National Bureau of Standards), Chapter 5.1}.

\bibitem{pccp} Metzler R, Jeon JH, Cherstvy AG, and Barkai E 2014
\textit{Phys. Chem. Chem. Phys.} \textbf{16}, 24128.

\bibitem{Wilcox1967}{Wilcox R M 1967 \textit{J Math. Phys.} \textbf{8} 962}.

\end{thebibliography}
\end{document}